\documentclass[12pt,a4paper]{article}
\usepackage{color,graphics,amsmath,epsfig,rotating,cite}

\begin{document}

%--------------------------------------------------------------------------
\newcommand\beqn{\begin{eqnarray}}
\newcommand\eeqn{\end{eqnarray}}
\newcommand{\ra}{\rightarrow}

\def\tb{\tan\beta}

\def\x{\chi}
\def\ti{\tilde}
\def\nt{\tilde \x^0}
\def\ch{\tilde \x^+}
\def\chpm{\tilde \x^\pm}
\def\st{\tilde t}
\def\stau{\tilde\tau}
\def\noi{\noindent}
\def\mser{m_{\tilde e_R}}
\def\sigv{\langle\sigma v\rangle}
\def\staul{\tilde\tau_L}
\def\staur{\tilde\tau_R}
\def\stopl{\tilde t_L}
\def\stopr{\tilde t_R}
\def\micromegas{{\tt micrOMEGAs\,2.1}}
\def\micro{{\tt micrOMEGAs}}
\def\darksusy{{\tt DarkSUSY}}
\def\isared{{\tt IsaRED}}

\newcommand{\mnt}[1]   {m_{\tilde\x^0_{#1}}}
\newcommand{\mch}[1]   {m_{\tilde\x^\pm_{#1}}}
\newcommand{\msf}[1]   {m_{\tilde f_{#1}}}
\newcommand{\mst}[1]   {m_{\tilde t_{#1}}}
\newcommand{\mstau}[1] {m_{\tilde\tau_{#1}}}

\newcommand{\eq}[1]  {\mbox{eq.~(\ref{eq:#1})}}
\newcommand{\fig}[1] {Fig.~\ref{fig:#1}}
\newcommand{\Fig}[1] {Figure~\ref{fig:#1}}
\newcommand{\tab}[1] {Table~\ref{tab:#1}}
\newcommand{\Tab}[1] {Table~\ref{tab:#1}}
\newcommand{\sect}[1] {Sect.~\ref{sec:#1}}

\newcommand{\ablabels}[3]{
  \begin{picture}(100,0)\setlength{\unitlength}{1mm}
    \put(#1,#3){\bf (a)}
    \put(#2,#3){\bf (b)}
  \end{picture}\\[-8mm]
}

\newcommand{\cdlabels}[3]{
  \begin{picture}(100,0)\setlength{\unitlength}{1mm}
    \put(#1,#3){\bf (c)}
    \put(#2,#3){\bf (d)}
  \end{picture}\\[-8mm]
}

\newcommand{\gsim}{\;\raisebox{-0.9ex}
           {$\textstyle\stackrel{\textstyle >}{\sim}$}\;}
\newcommand{\lsim}{\;\raisebox{-0.9ex}{$\textstyle\stackrel{\textstyle<}
           {\sim}$}\;}

%--------------------------------------------------------------------------

\vspace*{-18mm}
\begin{flushright}
  BONN-TH-2008-04\\
%  DESY ..........\\
  LAPTH-1239/08\\
  LPSC 0824\\
\end{flushright}
\vspace*{4mm}

\begin{center}

{\Large\bf
   Neutralino relic density from ILC measurements \\[3mm]
   in the CPV MSSM}\\[6mm]

{\large
   G. B\'elanger$^1$, O. Kittel$^2$, S. Kraml$^3$, H.-U.~Martyn$^{4,5}$, A. Pukhov$^6$}\\[6mm]

{\it
1) Laboratoire d'Annecy-le-Vieux de Physique Th\'eorique LAPTH, CNRS, Univ. de Savoie, 
 B.P. 110, F-74941 Annecy-le-Vieux Cedex, France\\
2) Physikalisches Institut der Universit\"at Bonn, Nussallee 12, D-53115 Bonn, Germany\\
3) Laboratoire de Physique Subatomique et de Cosmologie, UJF, CNRS/IN2P3, INPG,\\
    53 Avenue des Martyrs, F-38026 Grenoble, France\\
4) I. Physikalisches Institut, RWTH Aachen, Sommerfeldstrasse 14,\\ D--52074 Aachen, Germany\\
5) Deutsches Elektronen-Synchrotron DESY, Notkestr.\,85, D--22603 Hamburg, Germany\\
6) Skobeltsyn Inst. of Nuclear Physics, Moscow State Univ., Moscow 119992, Russia
}\\[6mm]

\end{center}

\begin{abstract}
We discuss ILC measurements for a specific MSSM scenario with CP phases, 
where the lightest neutralino is a good candidate for dark matter, annihilating  
efficiently through t-channel exchange of light staus. These prospective (CP-even) ILC 
measurements are then used to fit the underlying model parameters. A collider prediction 
of the relic density of the neutralino from this fit gives $0.116<\Omega h^2<0.19$ at 95\% CL. 
CP-odd observables, while being a direct signal of CP violation, do not help in further constraining 
$\Omega h^2$. The interplay with (in)direct detection of dark matter and with measurements of 
electric dipole moments is also discussed. Finally we comment on collider measurements at higher 
energies for refining the prediction of $\Omega h^2$.
\end{abstract}

%---------------------------------------------------------------------

\tableofcontents

%---------------------------------------------------------------------
\section{Introduction}
%---------------------------------------------------------------------

One of the prime motivations for a high-energy and high-luminosity
$e^+e^-$ linear collider is the possibility to do precision
measurements of new particles beyond the Standard Model. The
lightest of these new particles is often stable by virtue of a new
discrete symmetry and hence a candidate for the dark matter (DM) of the
Universe. One therefore hopes to be able to precisely determine
the properties of the DM in the laboratory, and in
particular to make a ``collider prediction" of its relic abundance,
which can be tested against cosmological models. For such a
collider prediction to be of interest, it must be as precise
as the value obtained from cosmological observations. This means a
precision of about 10\% (for 95\% CL)
to match WMAP+SDSS~\cite{Spergel:2006hy,Tegmark:2003ud,Dunkley:2008ie} 
or few percent to match expectations at the PLANCK satellite \cite{planck:2006uk}.
Moreover, if the annihilation cross section of the DM candidate is known, 
one can also predict direct and indirect DM detection cross sections 
as functions of astrophysical quantities such as galactic densities, 
see e.g.~\cite{Bertone:2004pz,Hooper:2008sn}.  

The possibility to make collider predictions of the cross sections
for annihilation of dark matter candidates has been examined
within specific supersymmetric scenarios. 
Within the constrained minimal supersymmetric standard model
(CMSSM), which has only a handful of free parameters, it has been
shown in a case study~\cite{Polesello:2004qy} that the LHC could
make a prediction for the DM relic density of the order of 10\%,
assuming the standard cosmological scenario; the ILC could reach a
precision of few percent~\cite{Bambade:2004tq} in several scenarios. 
The particular case of stau co-annihilation in the CMSSM 
was investigated for the LHC in \cite{Arnowitt:2008bz} 
and for the ILC in \cite{Martyn:2004jc} with similar conclusions.
In  the general MSSM, it was shown that the LHC
might match roughly the WMAP+SDSS precision in a favourable
scenario~\cite{Nojiri:2005ph} while the ILC could achieve much better
precision~\cite{Baltz:2006fm}. These conclusions, however, depend
very strongly on the scenario considered; many remain challenging
for both the LHC and the ILC~\cite{Baltz:2006fm}, see also \cite{Berger:2007ut}.
Moreover, the studies mentioned here mainly concentrated on a few 
scenarios that provide the correct amount of neutralino annihilation,
consistent with the WMAP+SDSS range. Furthermore --and more
importantly-- they assumed that CP is conserved, although
CP-violating (CPV) phases are generic in the MSSM.

It is well known that CPV MSSM phases can have an important
effect on the neutralino annihilation cross
sections~\cite{Falk:1995fk,Gondolo:1999gu,Nihei:2005va,Gomez:2005nr,Belanger:2006qa}.
The consequences of phases for direct and indirect detection were examined
in~\cite{Gondolo:1999gu,Nihei:2005va,Chattopadhyay:1998wb,Choi:2000kh,Nihei:2004bc}.
In \cite{Belanger:2006qa}, we
performed a comprehensive analysis of the impact of CP phases on
the relic density of neutralino dark matter, taking into account
consistently phases in all (co-)annihilation channels, and
carefully disentangling CPV effects due to modifications in the
couplings from pure kinematic effects.  We found variations in
$\Omega h^2$ solely from modifications in the couplings of up to
an order of magnitude. We concluded that the determination
of the relevant couplings (including CPV phases!) can be as
important for the prediction of $\Omega h^2$ as precision
measurements of masses, i.e.\ pure sparticle spectroscopy.

In this paper, we therefore consider a particular scenario of the
CPV MSSM, taken from \cite{Belanger:2006qa}, and investigate 
i)~which measurements are possible at the ILC, ii)~to which
precision the underlying MSSM parameters and the neutralino relic
density can be inferred from these measurements, and iii)~what 
are the implications for direct and indirect DM detection and measurements 
of electric dipole moments (EDMs).

Our scenario has
light gauginos and staus with masses below 200 GeV. The lightest
supersymmetric particle (LSP) is the lightest neutralino with a
mass of $\sim$80 GeV; it is dominantly bino. The two staus have
masses of about 100 GeV and 180 GeV, with a strong mixing between
the left- and right-chiral states. The neutralino LSP annihilates 
predominantly into tau pairs, with the annihilation cross section
being sensitive to the stau mixing. Although the
$\stau_1$ is very light, the scenario does not rely on
coannihilation but on t-channel stau exchange. We therefore refer
to it as the ``stau-bulk'' scenario. Such a scenario occurs in the
CP-conserving MSSM only for $M_1<0$.

We further impose that the sfermions of the first and second generation are heavy
in order to avoid the strong EDM constraints. %the electric dipole moments (EDMs).
The resulting mass pattern, light staus but TeV-scale selectrons and smuons,
is not found in SUSY models where universality among scalar masses is imposed
at a high scale.
It is important to note that our scenario, despite having several particles below 200~GeV,
is quite challenging for colliders. At the LHC, SUSY events are dominated by squark and
gluino production followed by cascade decays leading to jets plus $\tau$'s plus $E_T^{\rm miss}$.
At the ILC, two neutralinos, the lighter chargino, and the two staus are within kinematic reach
with $\sqrt{s}=500$~GeV. However, production of these sparticles again only leads to
taus plus missing energy.

The paper is organised as follows. Section~2 describes the
detailed setup of our ``stau-bulk'' scenario. Expectations for ILC
measurements are given in Section~3. In Section~4, we present our
results concerning the determination of the model parameters and
the prediction for the relic density of dark matter. In Section~5 we 
then discuss CP-sensitive observables, and in Section~6 
additional possibilities to test and constrain the
model at higher energies, through EDM measurements or through
direct DM detection experiments. Finally, Section~7 contains our
conclusions.

%--------------------------------------------------------------------------------------------------------------
\section{Setup of the stau-bulk scenario}
\label{sec:benchmark}
%--------------------------------------------------------------------------------------------------------------

In the MSSM, the parameters that can have CP phases are the
gaugino and Higgsino mass parameters, $M_i=|M_i|e^{i \phi_i}$, with $i=1,2,3$,
and $\mu=|\mu|e^{i \phi_\mu}$, and the trilinear sfermion-Higgs
couplings, $A_f=|A_f|e^{i \phi_f}$. Not all of these phases are, however,
physical. The physical combinations are Arg$(M_i\mu)$ and Arg$(A_f\mu)$.
Allowing for CP-violating phases, in particular non-vanishing $\phi_f$,
induces a mixing between the two CP-even states $h^0$, $H^0$ and the
CP-odd state $A^0$.  The resulting mass eigenstates $h_1$, $h_2$, $h_3$
($m_{h_1}<m_{h_2}<m_{h_3}$) are no longer eigenstates of CP.
Therefore the charged Higgs mass, $m_{H^+}$, is typically used as input
parameter in the CPV-MSSM.

In this paper we investigate the ``stau bulk'' region of \cite{Belanger:2006qa},
which appears for light staus and large phase of $M_1$. The point is the
following.
In the conventional case, there are mainly two mechanisms that can make a light
bino-LSP annihilate efficiently enough: resonant annihilation through the light Higgs,
or co-annihilation with a sparticle that is close in mass --in the CMSSM typically
the lighter stau. The so-called ``bulk region"  where the LSP annihilates through
t-channel exchange of very light sleptons is largely excluded by LEP.
For large phases of $M_1$ we have found, however, that the couplings of the
neutralino to staus can be sufficiently enhanced such that a new region opens up,
where the $\nt_1$--$\stau_1^{}$ mass difference is too large for co-annihilation
to be efficient but the LSP annihilates into taus through t-channel exchange of
both $\stau_1$ and $\stau_2$.

To define a benchmark point for such a scenario, we choose the following
input parameters at the electroweak scale:
\begin{equation}
\begin{array}{llll}
  M_1 = 80.47~{\rm GeV},\quad  & M_2 = 170.35~{\rm GeV},\quad & M_3 = 700~{\rm GeV},\quad
  &  \phi_1 = 180^\circ, \\
  M_{\tilde L_3} = 138.7~{\rm GeV},  & M_{\tilde E_3} = 135.2~{\rm GeV},  & A_\tau= 60~{\rm GeV},
  & \phi_\tau = 0,\\
  \mu = 600~{\rm GeV},  &  \tan\beta = 10, &  \phi_\mu = 0.
%  M_{\tilde Q_3} = M_{\tilde U_3} = M_{\tilde D_3} & = A_{t,b} = 1~{\rm TeV}.
\end{array}
\label{eq:defpar}
\end{equation}
All other parameters, i.e.\ sfermion masses, $A_{t,b}$, and $m_{H^+}$ are set to 1~TeV for simplicity.
This way EDM constraints are avoided when varying $\phi_1$ and $A_\tau$; all other phases
are assumed to vanish.

We use {\tt CPsuperH}~\cite{Lee:2003nt} to compute Higgs and
sparticle masses and mixing angles, and
\micromegas~\cite{Belanger:2006is,Belanger:2008} to compute the relic density,
EDMs, and (in)direct detection cross sections. The mixing angle in the
stau sector writes
\begin{eqnarray}
\label{eq:mixing}
  \tan 2\,\theta_{\stau}&=&\frac{-2\,m_\tau\,(A_\tau-\mu\tan\beta)}
           {m^2_{\stau_R} - m^2_{\stau_L}}
\end{eqnarray}
and for our scenario is completely dominated by the $\mu\tan\beta$
term.
 The mass spectrum resulting from \eq{defpar}  is
given in \tab{defmass}. The particles accessible at ILC with
$\sqrt{s}=500$~GeV are $\nt_1$, $\nt_2$, $\ch_1$, $\stau_1$,
$\stau_2$ and $\tilde{\nu}_\tau$. Production cross sections and
branching ratios are discussed in Section~3. At this stage we just
note that the scenario is rather challenging to resolve
experimentally because SUSY events involve only $\tau$'s and
missing energy.

\begin{table*}[tb]
\begin{center}
\begin{tabular}{|l|l|l|l|l|l|l|l|l|}
\hline\hline
   & $\nt_1$ & $\nt_2$ & $\nt_3$ & $\nt_4$ & $\ch_1$ & $\ch_2$& $h_1$ & $h_{2,3}$\\ \hline
   & 80.7 & 164.9  & 604.8     & 610.5  & 164.9& 612.1 & 116.1 & 997.  \\
\hline\hline
  & $\tilde{\tau}$ & $\tilde{\nu}_\tau$ & $\tilde{e}$ & $\tilde{\nu}_e$
  & $\tilde{u}$ & $\tilde{d}$ & $\tilde{t}$ & $\tilde{b}$\\ \hline
R(1)   &  100.9  &  --  & 1000.9  & --  & 999.4 &    1000.3 & 939.1 & 995.6 \\ \hline
 L(2)   & 177.2   &  123.1 & 1001.1  & 998.0 & 998.6 & 1001.7 & 1075.6 & 1006.4 \\
\hline\hline
\end{tabular}
\caption{Masses of particles for the input parameters of \eq{defpar}. } \label{tab:defmass}
\end{center}
\end{table*}

The relic density of the $\nt_1$ is $\Omega h^2=0.130$ at the nominal point \eq{defpar}.
As mentioned above, the dominant annihilation channel is
into tau pairs (more than 95\% of the total contribution) through
t-channel exchange of staus. The contribution of both
$\tilde{\tau}_1$ and $\tilde{\tau}_2$ is crucial in bringing $\Omega h^2$ to the
desired range, $0.094 < \Omega h^2 < 0.136$~\cite{Hamann:2006pf}.
Indeed, for $\phi_1=0$ (or $M_1$ positive) one would have $\Omega h^2=0.167$.
Note also that the mass splitting between the stau-NLSP and the LSP is too large
for any significant contribution from co-annihilation.

\begin{figure}[t]
\centerline{\epsfig{file=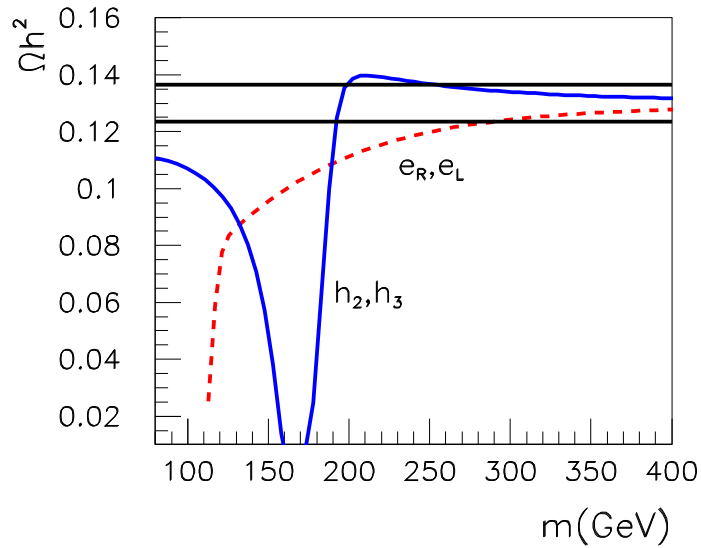, width=10cm}} \vspace{-5mm}
\caption{Dependence of $\Omega h^2$ on $m_{\tilde e,\tilde\mu}$
(dashed red line) and
     $m_{h_{2,3}}$ (full blue line) for the nominal point \eq{defpar}.
     The horizontal lines indicate $\delta\Omega/\Omega=\pm5\%$.}
 \label{fig:varomega}
\end{figure}

The precision with which $\Omega h^2$ can be inferred from ILC measurements therefore
depends not only on the accuracy of the sparticle spectroscopy (mass measurements)
but on the determination of all parameters of the neutralino sector
($M_1$, $M_2$, $\mu$, $\tan\beta$, $\phi_1$), which determine the LSP mass
and couplings, and the four parameters of the stau
sector ($M_{\tilde\tau_L}$, $M_{\tilde\tau_R}$, $A_\tau$, $\phi_\tau$).
The dependence on $\phi_1$, and to a much lesser extent  $\phi_\tau$, 
originates from the $\nt_1\tilde{\tau}_{1,2}^{}\tau$ couplings.
In addition, particles which are too heavy to be produced at ILC could
have some influence. Indeed, the exclusion of selectrons and smuons
up the the kinematic limit, $m_{\tilde l}>250$~GeV, leaves an uncertainty
in $\Omega h^2$ of about 7\%. The influence of heavy Higgs states
gives $\delta\Omega /\Omega\simeq 5\%$ for $m_{h_{2,3}}\simeq 250$~GeV.
The combined effect from sleptons and Higgses is actually smaller, because their
contributions work against each other, so we have $\delta\Omega /\Omega\simeq 7\%$
as overall uncertainty from the unknown part of the spectrum.\footnote{Notice, however, 
that the measurement of the total $e^+e^-\to \tau^+\tau^-$ cross section above 
the $\ti\x^+_1\ti\x^-_1$ threshold, being dominated by chargino-pair production, provides an 
indirect constraint on the $\ti\nu_e$ mass, and hence also on that of $\ti e_L$, of 
$m_{\tilde\nu_e,\ti e_L}\gsim 900$~GeV.}
The dependence on $m_{\tilde e,\tilde\mu}$ and $m_{h_{2,3}}$ is illustrated in \fig{varomega}.

%%%%%%%%%%%%%%%%%%%%%%%%%%%%%%%%%%%%%%%%%%%%%%%%%%
%%%%%%%%%%%%%%%%%%%%%%%%%%%%%%%%%%%%%%%%%%%%%%%%%%

We assume here that the mass scale of the squarks and gluino is known from
LHC. The LHC can also provide additional information on the heavy
Higgs states: a charged Higgs with mass up to $m_{H^+}\simeq
500$~GeV could be discovered/excluded at the LHC in the channel
$gb\to H^-t$ with $H^-\to\tau\nu$ for
$\tan\beta=10$\cite{atlas:1999fq}. If the neutral heavy Higgses
weigh more than $250$~GeV, the LHC has limited discovery potential
using SM decay modes at intermediate values of $\tan\beta$.
However, a recent analysis \cite{Bisset:2007mi} has shown how to
exploit the Higgs into chargino/neutralino decay modes into four
leptons (electrons or muons) in this region. Although no dedicated
analysis has been performed for decays into taus, a similar
coverage as in \cite{Bisset:2007mi} can be expected in our
scenario~\cite{filip}.

{

%---------------------------------------------------------------------------------------------------------
%------------------------------------------------------------------------------------------------------------------ 
\section{ILC measurements} 
%------------------------------------------------------------------------------------------------------------------ 
 
\newcommand{\hdick}{\noalign{\hrule height1.4pt}} 
\newcommand{\eV}  {\mathrm{eV}} 
\newcommand{\MeV} {\mathrm{MeV}} 
\newcommand{\GeV} {\mathrm{GeV}} 
\newcommand{\TeV} {\mathrm{TeV}} 
\newcommand{\fb}  {\mathrm{fb}} 
\newcommand{\fbi} {\mathrm{fb}^{-1}} 
\newcommand{\ab}  {\mathrm{ab}} 
\newcommand{\abi} {\mathrm{ab}^{-1}}  
 
\newcommand{\sek} {\mathrm{s}} 
\newcommand{\cL } {{\cal L}} 
\newcommand{\cP } {{\cal P}} 
\newcommand{\cB } {{\cal B}} 
 
\def\ee{e^+e^-} 
\def\ti    {\tilde} 
\def\sf    {{\ti f}} 
\def\sq    {{\ti q}} 
\def\str   {{\ti t}_R} 
\def\stl   {{\ti t}_L} 
\def\st    {{\ti t}} 
\def\sb    {{\ti b}} 
\def\tb    {\bar{\ti t}} 
\def\bb    {\bar{\ti b}} 
\def\stau  {{\ti\tau}} 
\def\staum {{\ti\tau}^-} 
\def\staup {{\ti\tau}^+} 
\def\snu   {{\ti\nu}} 
\def\sell  {{\ti\ell}} 
\def\sll   {{\ti\ell}} 
\def\sl    {{\ti\ell}} 
\def\cx    {\ti {\chi}} 
\def\ch    {\ti {\chi}} 
\def\cp    {\ti {\chi}^+} 
\def\cm    {\ti {\chi}^-} 
\def\cone  {\ti \chi^-_1} 
\def\ctwo  {\ti \chi^-_2} 
\def\nt    {\ti {\chi}^0} 
\def\none  {\ti \chi^0_1} 
\def\ntwo  {\ti \chi^0_2} 
\def\nthre {\ti \chi^0_3} 
\def\nfour {\ti \chi^0_4} 
\def\sg    {\ti g} 
\def\sG    {\ti G} 
\def\sq    {\ti q} 
 
\def\cth   {\cos\theta} 
\def\sth   {\sin\theta} 
\def\tsf   {\theta_{\ti f}} 
\def\tst   {\theta_{\ti t}} 
\def\tsb   {\theta_{\ti b}} 
\def\tstau {\theta_{\ti\tau}} 
\def\csf   {\cos\theta_{\ti f}} 
\def\cst   {\cos\theta_{\ti t}} 
\def\csb   {\cos\theta_{\ti b}} 
\def\cstau {\cos\theta_{\ti\tau}} 
 
\def \Eslash {E^{\rm miss}}  %{E \kern-.75em\slash } 
 
\subsection{Event generation and analysis} 
 
The analysis presented in this section assumes a detector as 
described in the {\sc Tesla} TDR~\cite{AguilarSaavedra:2001rg}. 
The simulations are based on {\sc Simdet}~4.02~\cite{simdet} 
with an acceptance $\theta > 125$~mrad and 
$e,\,\gamma$ veto of $\theta > 4.6$~mrad. 
Background and signal events  are generated with 
{\sc Pythia}~6.2~\cite{Sjostrand:2000wi} including 
beam polarisations $(\cP_{e^-},\cP_{e^+}) = (0.8, 0.6)$, 
QED radiation and  beamstrahlung \`a la {\sc Circe}~\cite{circe}. 
$\tau$ decays and polarisation are treated with {\sc Tauola}~\cite{Jadach:1993hs}. 
The SM background includes $\ee \to W^+ W^-$  
and the negligible pair production $\ee\to \tau^+\tau^- (\gamma)$. 
The $\gamma\gamma$ background is also negligible with 
$\sigma(\ee\to e^+e^-\tau^+\tau^-) = 4.5\cdot 10^5~\fb$ and an 
acceptance $< {\rm few} 10^{-6}$. 
  
To measure the particle masses, two methods are available:
threshold scans and endpoint methods.  Both will be useful in our 
case. In slepton decays to LSP, $\sell^- \to \ell^- \, \nt_1$, the 
endpoints, i.e. maxima and minima $E_{+/-}$ in the flat lepton energy 
spectra:  
\begin{eqnarray} 
  E_{+/-} & = & 
        \frac{\sqrt{s}}{4} 
        \left ( 1 - \frac{m_{\cx}^2}{m_{\sl}^2} \right ) 
        \, (1 \pm \beta)   \; , 
          \label{eq:endpoint} 
\end{eqnarray} 
where $\beta = \sqrt{1-4m_{\sl}^2/s}$, can be 
used to extract the slepton and neutralino mass. 
 
The mixing angle in the stau sector, eq.~(\ref{eq:mixing}), can 
either be determined from measurements of the polarized cross 
section or from the measurement of the $\tau$ polarisation 
$\cP_\tau$ in the decay 
$\stau_1 \to \tau \nt_1$~\cite{Nojiri:1994it,Nojiri:1996fp,Boos:2003vf}. 
Note that  $\cP_\tau$ depends not only on the mixing anlge $\tstau$ of the 
staus but also on the mixing of the lightest neutralino. 
 
The selection criteria used in $e^+e^-\to\stau^+_1\stau^-_1 \to\tau^+\tau^-\Eslash$ 
are essentially to require two acoplanar $\tau$ jets ($\Delta\phi<180^\circ$ and 
$m_\tau<2\;\GeV$) being produced centrally ($|\!\cos\theta| < 0.75$). 
In addition, topology dependent $p_\perp$ cuts are applied to reject
$\gamma\gamma$ reactions~\cite{Martyn:2004jc}. 
The overall efficiency for $\stau_1$-pair reconstruction is $\epsilon \simeq 0.32$
at $\sqrt{s} = 280\;\GeV$ and varies only slowly with the cms energy.
The same selection criteria are applied to detect the other sparticle decays
during an energy scan. 

Despite the large statistics, the leptonic $\tau$ decays are not useful due
to the large $WW$ background. We therefore only analyse the decay modes 
$\tau\to\pi\nu_\tau,\;\rho\nu_\tau,\;3\pi\nu_\tau$ with branching 
ratios 
$\cB(\tau\to\pi\nu_\tau) = 0.111, \ 
\cB(\tau\to\rho\nu_\tau) = 0.254, \ 
\cB(\tau\to3\pi\nu_\tau) = 0.194$.

\subsection{Masses from threshold scan} 
 
At $\sqrt{s}=500\;\GeV$, all SUSY signals are completely dominated by 
the $\tau\tau\Eslash$~ topology. The challenge is then to 
disentangle the various sources. The strategy we propose is to 
scan downwards in energy in steps of 10 GeV in order to find 
thresholds. This is done  using different beam polarisations 
$e^-_Re^+_L$ [$\sigma_{RL}$] and $e^-_Le^+_R$ [$\sigma_{LR}$] 
where $\cP_{e-}=\pm0.8$ and $\cP_{e^+}=\mp0.6$. 
The $\stau_i^+\stau_j^-$ pair production is best detected in the $\sigma_{RL}$
polarisation mode,
while $\sigma_{LR}$ is dominated by $\cp_1\cm_1$ production 
The results of a scan of the visible cross 
sections $\ee\to\tau\tau\Eslash$ \ are shown in \fig{cmscan}. 
Clearly, integrated luminosities of $2\;\fbi$ per step in energy are sufficient. 
Fitting the excitation curves, 
$\sigma_{\stau\stau}\sim \beta^3$  of \fig{cmscan}a
and $\sigma_{\cx^+\cx^-}\sim\beta$ of \fig{cmscan}b,
allows a determination of the sparticle masses.
Note that the mixed $\stau_1\stau_2$ production is hardly detectable and can
only be accomodated in a global fit including all channels.
The results of a fit for $\stau_1$, $\cp_1$ and $\stau_2$ masses together with
the total integrated luminosity used in each case are listed in 
Table~\ref{tab:scan_fit}. 
The expected accuracies can easily be scaled with the luminosity.

\begin{figure} \centering 
  \epsfig{file=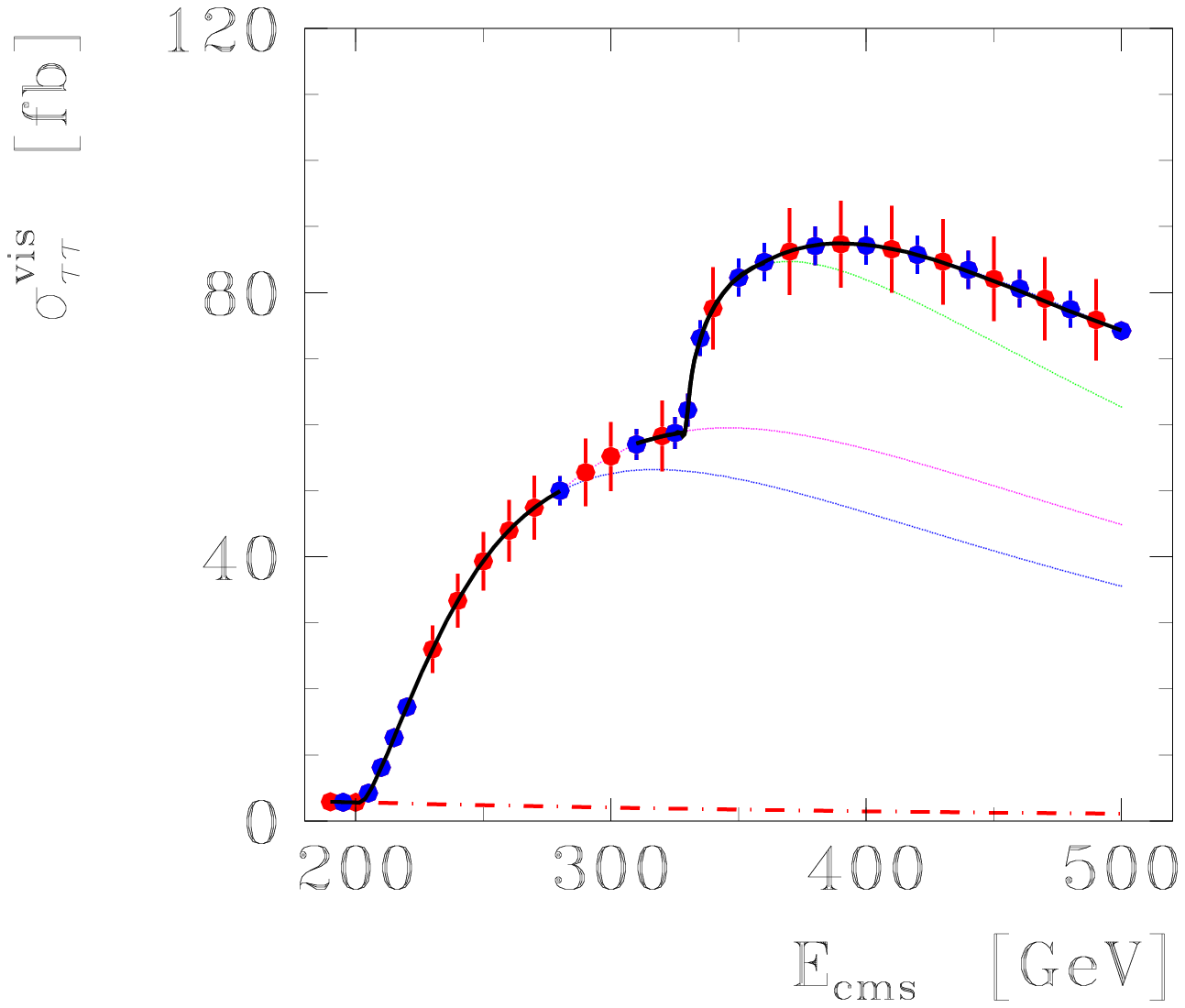,angle=90,width=.48\textwidth} \hfill
  \epsfig{file=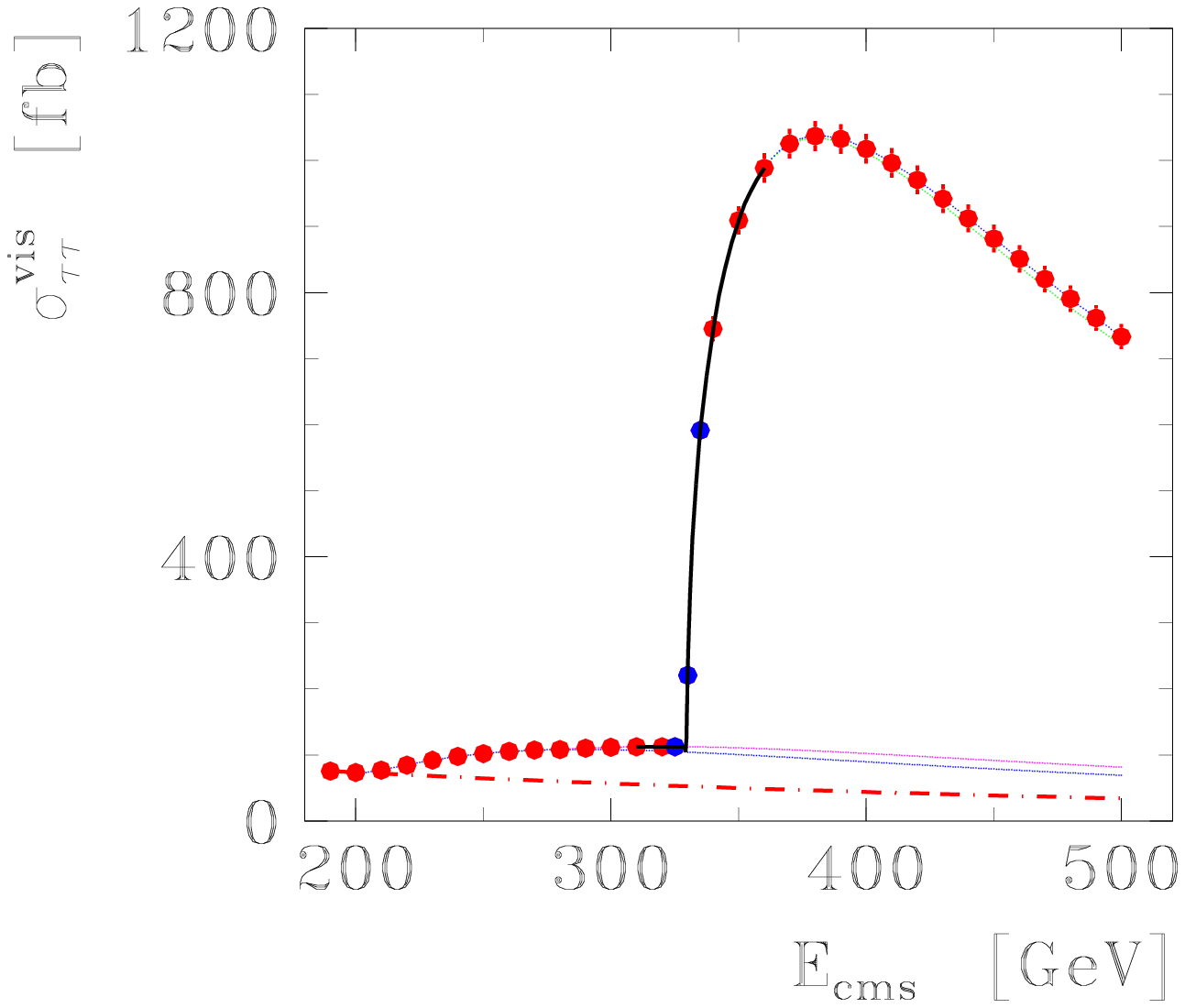,angle=90,width=.48\textwidth} 
  \caption{Visible cross section as a function of cms energy 
    of the inclusive reaction 
    $e^+_L e^-_R\to \tau^+ \tau^-  \,E^{\rm miss}  $ (left) 
    and 
    $e^+_R e^-_L\to \tau^+ \tau^-  \,E^{\rm miss}  $ (right). 
    The red (blue) points correspond to integrated luminosities 
    of $\cL = 2\,\fbi (10\,\fbi)$, 
    the contributions of $W^+W^-$ (dash-dot), 
    $\stau_1^+\stau_1^-$ (blue), 
    $\stau_1^\pm\stau_2^\mp$ (pink), $\cx^+_1\cx^-_1$(green) and 
    $\stau_2^+\stau_2^-$ are indicated. 
  } 
  \label{fig:cmscan} 
\end{figure} 
 
\begin{table}[h] 
\begin{center} 
    \begin{tabular}{l c c} 
      reaction              & $m~[\GeV]$ & $\cL~[\fbi]$ \\ \hdick 
      $e^+_Le^-_R\to\stau_1\stau_1$    & $100.92 \pm 0.40$  & $ 50 $ \\ 
      $e^+_Re^-_L\to\cp_1\cm_1$        & $164.88 \pm 0.015$ & $ 30 $ \\ 
      $e^+_Le^-_R\to\stau_2\stau_2$    & $188.2 \pm 9.1$    & $ 100 $ \\ 
    \end{tabular} 
  \caption{Results of mass fits from the excitation curves indicating the 
    reactions, beam polarisation and assumed integrated luminosity} 
\label{tab:scan_fit} 
\end{center} 
\end{table}

\subsection{LSP mass from stau decays} 

An alternative method for measuring the $\stau_1$ mass 
and/or the $\nt_1$ LSP mass
is to use the energy spectra in $\stau_1\to \tau\nt_1$ decays. 
However, the $\tau$ energy spectrum reconstructed from the $\tau$ decay 
products is no longer flat due to undetected neutrinos, and 
the extraction of the endpoint energies, see eq.~(\ref{eq:endpoint}), 
gets more involved~\cite{Nojiri:1996fp,Boos:2003vf}.
The upper endpoint energy $E_+$ can be identified with the maximum energy of
the decay products and can still be determined rather well.
The lower endpoint energy $E_-$ is in general completely distorted.
It may be reconstructed from the pion energy $E_\pi$ spectrum of
$\tau \to \pi\; \nu_\tau$, but the expected precision is poor due
the low branching ratio 
and a polarisation $\cP_\tau$ dependent shape of the energy distribution.  
Therefore, determining both stau and neutralino masses from the energy spectra
alone is not very meaningful and we use the
additional information on $m_{\stau_1}$ from the threshold scan.

In order to study the properties of stau decay spectra, the maximal energy for
$\stau_1^+\stau_1^-$ production at $\sqrt{s}=280\;\GeV$
just below the threshold for other sparticles is chosen.
A fit to the $E_\pi$ spectrum (not shown) yields 
$E_- = 7.8 \pm 0.50\;\GeV$ and $E_+ =42.8 \pm 0.30\; \GeV$.
In $\tau \to \rho\; \nu_\tau \to \pi^\pm\pi^0\; \nu_\tau$ and 
$\tau \to 3 \pi\; \nu_\tau \to \pi^\pm\pi^\pm\pi^\mp\; \nu_\tau 
                           + \pi^\pm\pi^0\pi^0\; \nu_\tau $ 
the energies $E_\rho$ and $E_{3\pi}$  have the advantage of being 
independent of $\cP_\tau$. 
The analysis of $E_{jet} \ (= E_\rho + E_{3\pi})$ spectrum is shown in
\fig{espectra}, a fit to the energy distribution using an analytic 
formula gives the upper endpoint $E_+ =42.8 \pm 0.20\; \GeV$.
Applying eq.~(\ref{eq:endpoint}) together with the results from the excitation
curve leads to a LSP mass of
  $\mnt{1} =  80.67 \pm 0.35 \;\GeV$
and a stau mass of
  $m_{\stau_1} = 100.92 \pm 0.39 \;\GeV$.
The $m_{\nt_1} - m_{\stau_1}$ correlation 
is shown in \fig{espectra}. 
Obviously an improved accuracy on $m_{\stau_1}$ will result in a smaller error 
on the LSP mass.

\begin{figure} \centering 
  \epsfig{file=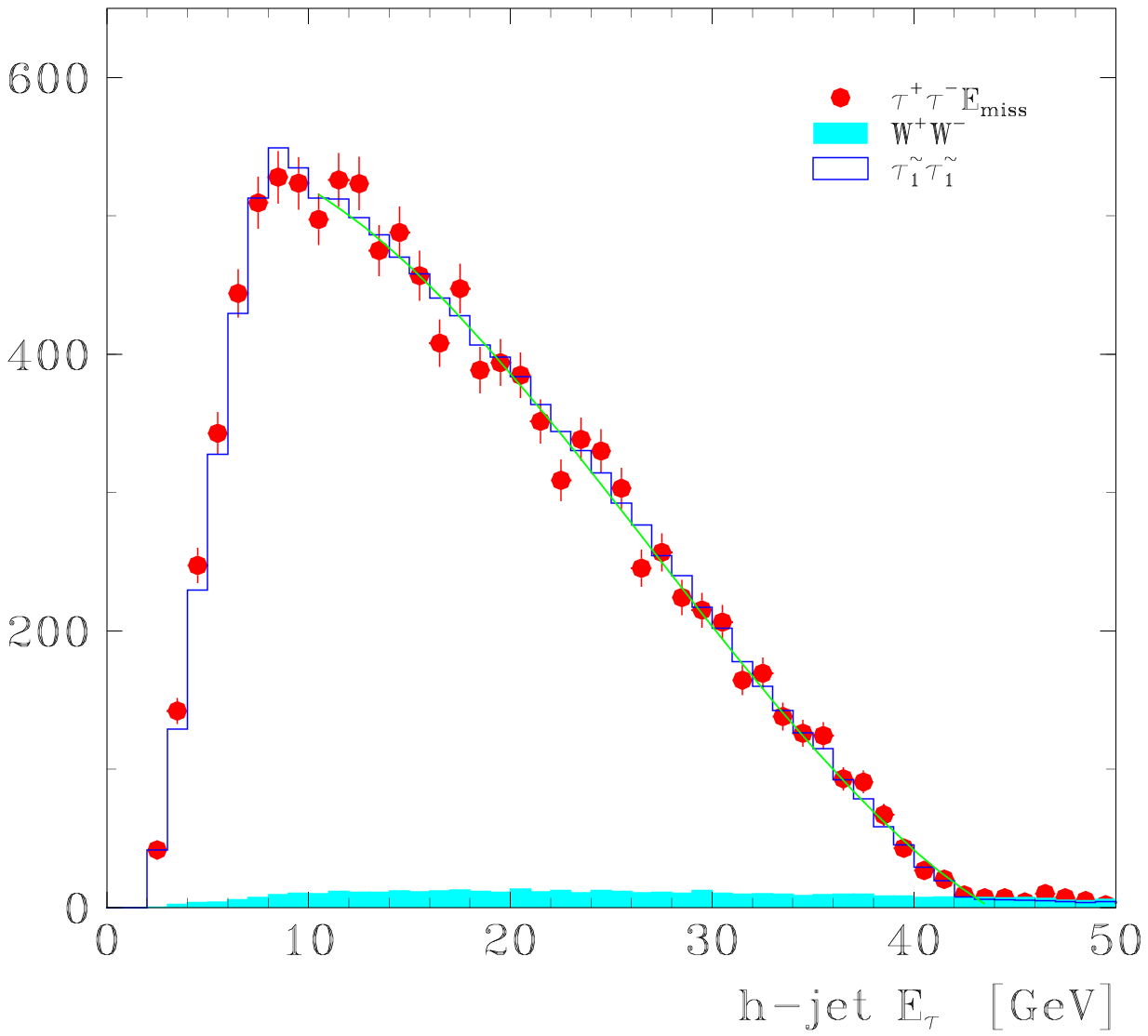,angle=90,width=.48\textwidth} \hfill
  \epsfig{file=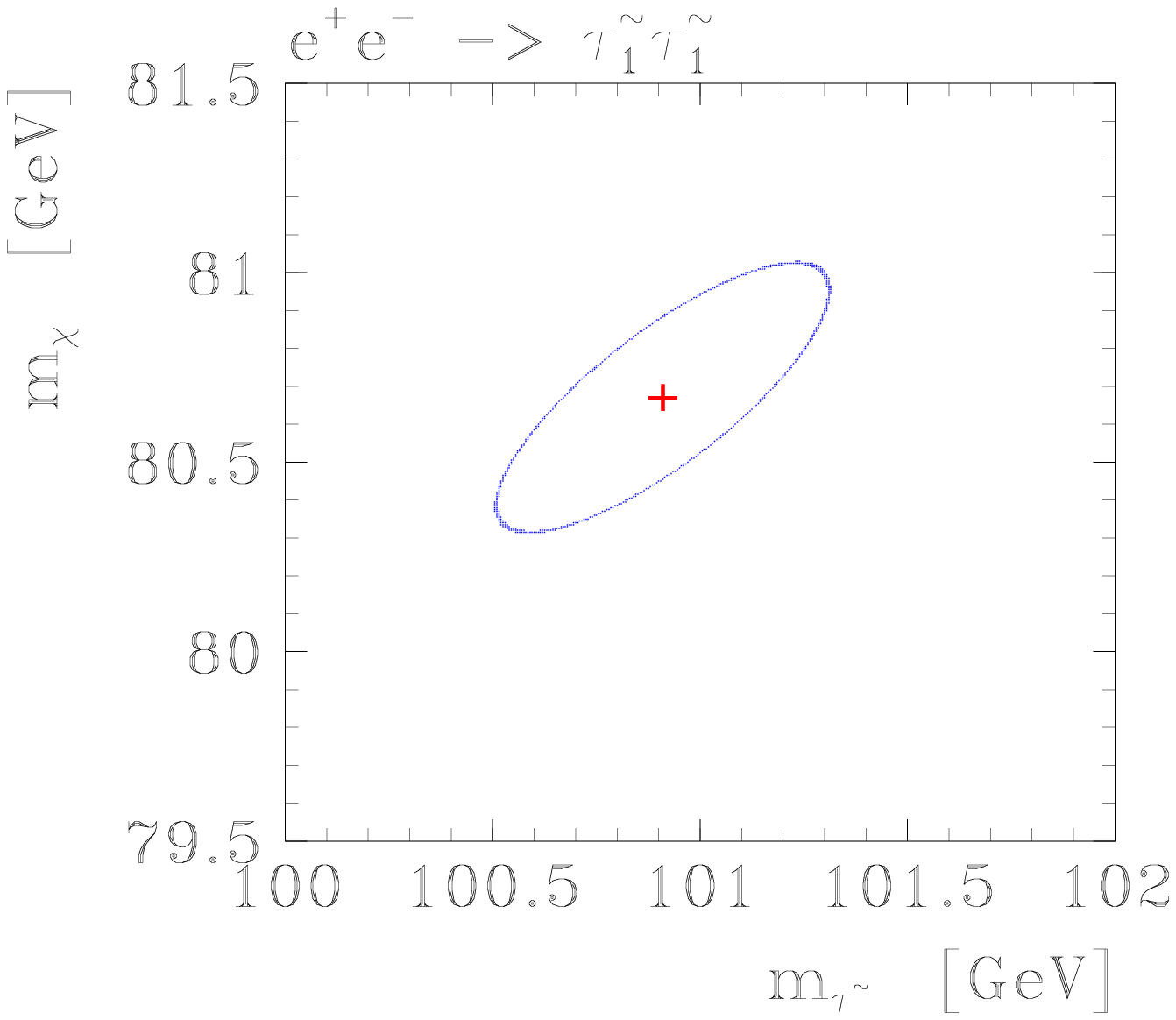,angle=90,width=.48\textwidth} 
  \caption{Energy spectrum $E_{jet}$ from 
    $\tau \to \rho\; \nu_\tau +3 \pi\; \nu_\tau$ %(left) 
    of the reaction $e^+_L e^-_R\to \stau^+_1 \stau^-_1$ 
    including background from $W^+W^-\to\tau^+ \tau^- \; E_{miss}$; 
    the data points represent a simulation assuming $\cL = 200\,\fbi$ at 
    $\sqrt{s}=280\,\GeV$, 
    the blue histogram corresponds to a very high statistics sample, 
    the green curve is a fit to an analytic formula (left).
    Correlation $m_{\nt_1}$ vs $m_{\stau_1}$ using 
    the stau mass from the excitation curve and the end point energy 
    from the $E_{jet}$ spectrum (right).} 
    \label{fig:espectra} 
\end{figure}

\subsection{Stau mixing angle} 
 
The stau mixing angle can be determined from measurements of 
polarized cross sections.  
Again, the energy for $\stau_1$ pair production is chosen below the threshold
for other sparticles production, that is $\sqrt{s}=280$~GeV. 
At this energy we expect $\sigma_{RL} = 155.5 \pm 2.2 \,\fb$ 
% and $\sigma_{LR} = 157.3 \pm 3.5 \,\fb$. 
% is less precise because of the large background from W pairs. 
where the error corresponds to an integrated luminosity of
$\cL=200\,\fbi$
and includes an estimate of systematics from acceptance calculations,
branching ratios and $WW$ background subtraction; 
for comparison see the visible cross section in \fig{cmscan}.

\begin{figure} \centering 
  \epsfig{file=cpv_mixingRL.eps,angle=90,width=.45\textwidth} \hfill
  \epsfig{file=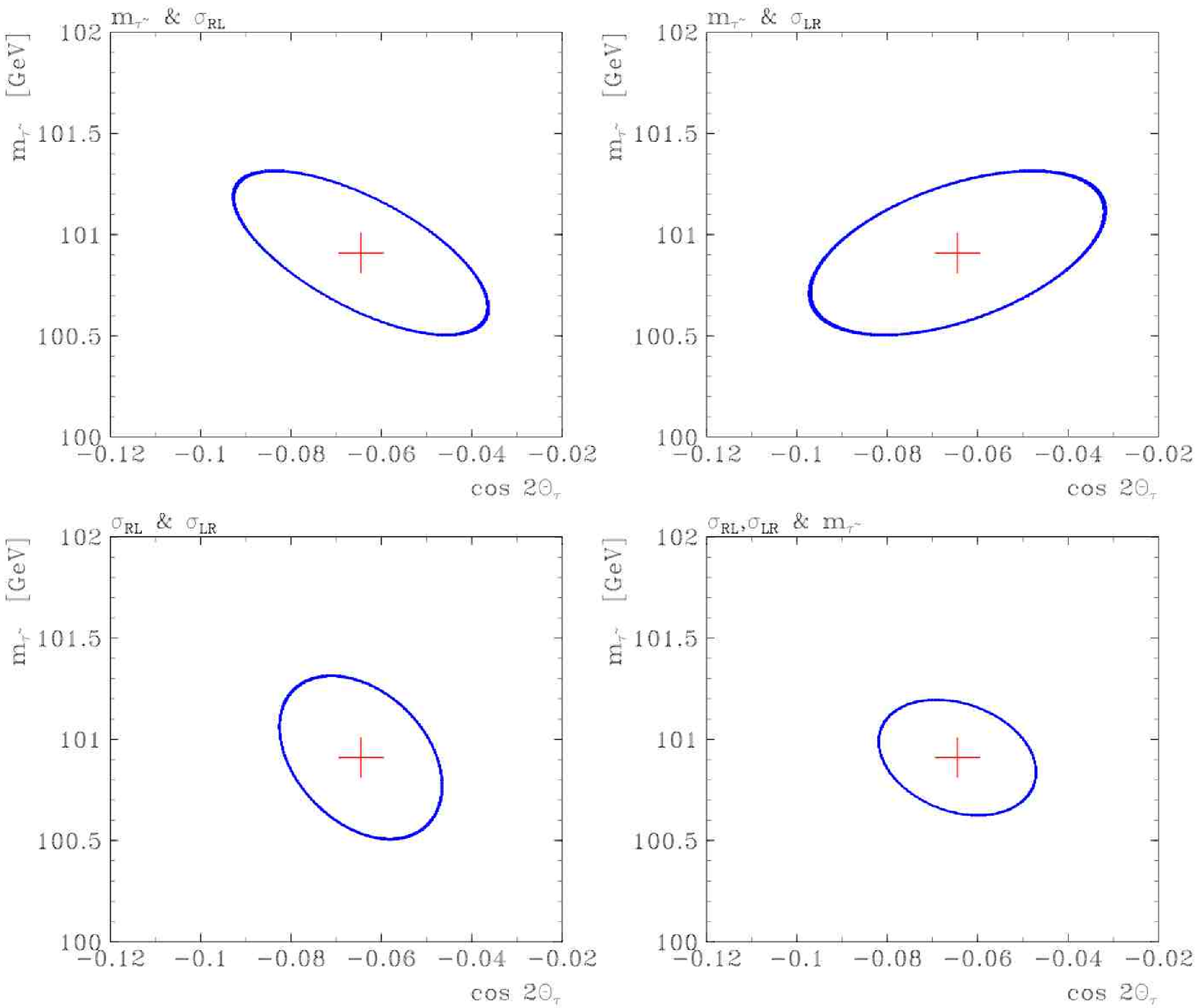,angle=90,width=.48\textwidth,
    bbllx=280pt,bburx=600pt,bblly=335pt,bbury=680pt,clip=} 
  \caption{Dependence of $\stau_1\stau_1$ polarised cross section 
    $\sigma_{RL}$  on the $\stau$ mixing angle 
    $\cos 2\tstau$ at $\sqrt{s}=280\,\GeV$, 
    the blue band indicates a measurement assuming $\cL = 200\,\fbi$
    (left):
    correlation $m_{\stau_1}$ vs $\cos 2\tstau$ using the stau mass from the
    excitaion curve the cross section $\sigma_{RL}$ (right).}
  \label{ctxsec} 
\end{figure}

The polarised cross section $\sigma_{RL}$
dependence on the mixing angle $\cos2\tstau$~\cite{Boos:2003vf} is 
shown in Fig.~\ref{ctxsec}.
Fitting both the mixing angle 
$\cos2\tstau$ and the mass $\mstau{1}$ to the cross sections in the continuum
and at threshold yields $\cos2\tstau = -0.065 \pm 0.028 $
and $m_{\stau_1} = 100.92 \pm 0.40\;\GeV$.
The $m_{\stau_1}$--$\cos2\tstau$ correlation contour is shown in
Fig.~\ref{ctxsec}.
The accuracy can be improved by adding another cross section measurement with
different polarisation, e.g. $\sigma_{LR} = 157.3 \pm 3.5 \,\fb$ assuming
$\cL=200\,\fbi$.
%Incidentally a measurement of the two polarised cross sections give the same
%stau mass precision as the threshold scan. 
Using all observables,
$\sigma_{RL}$, $\sigma_{LR}$ and $m_{\stau_1}(\sigma_{thr})$, one expects
a precision of $\delta m_{\stau_1} = 0.35\;\GeV$ and
$\delta\!\cos2\tstau = 0.017$.

\subsection{Tau polarisation} 
 
The polarisation $\cP_\tau$ of the tau stemming from the $\stau_i\to\tau\nt_1$ 
decay gives additional information on the stau and neutralino mixings \cite{Nojiri:1994it}. 
Since $\cos2\tstau$ is obtained from polarized cross sections as explained above, 
$\cP_\tau$ is in particular useful to constrain the gaugino--higgsino composition 
of the LSP. Here, we exploit the decay  $\tau\to\rho\nu_\tau\to \pi\pi^0\nu_\tau$. 
The ratio $z_\pi=E_\pi/E_\rho$ is indeed sensitive to $\cP_\tau$ and 
independent of  $m_{\stau_1}$. For a right-handed $\tau$ 
($\cP_\tau=+1$), the $\rho$ is longitudinally polarized and the 
distribution ${\rm d}\sigma/{\mathrm d} z_\pi \propto (2 z_\pi-1)^2$ 
is peaked near $z_\pi\to 0{\rm ~or~ }1$, while for a left-handed $\tau$ 
($\cP_\tau=-1$), the $\rho$ is transversely polarized and the 
distribution ${\rm d}\sigma/{\rm d} z_\pi \propto 2 
z_\pi(1-z_\pi)$ is peaked at $z_\pi=0.5$. 
The analysis of the $z_\pi$ spectrum is shown in 
Fig.~\ref{taupol}. The results of a fit to the $\stau_1$ 
polarisation leads to $\cP_\tau=0.64 \pm 0.035$ for an input value 
 $\cP_\tau=0.641$.  
 
\begin{figure}[htb] \centering 
  \epsfig{file=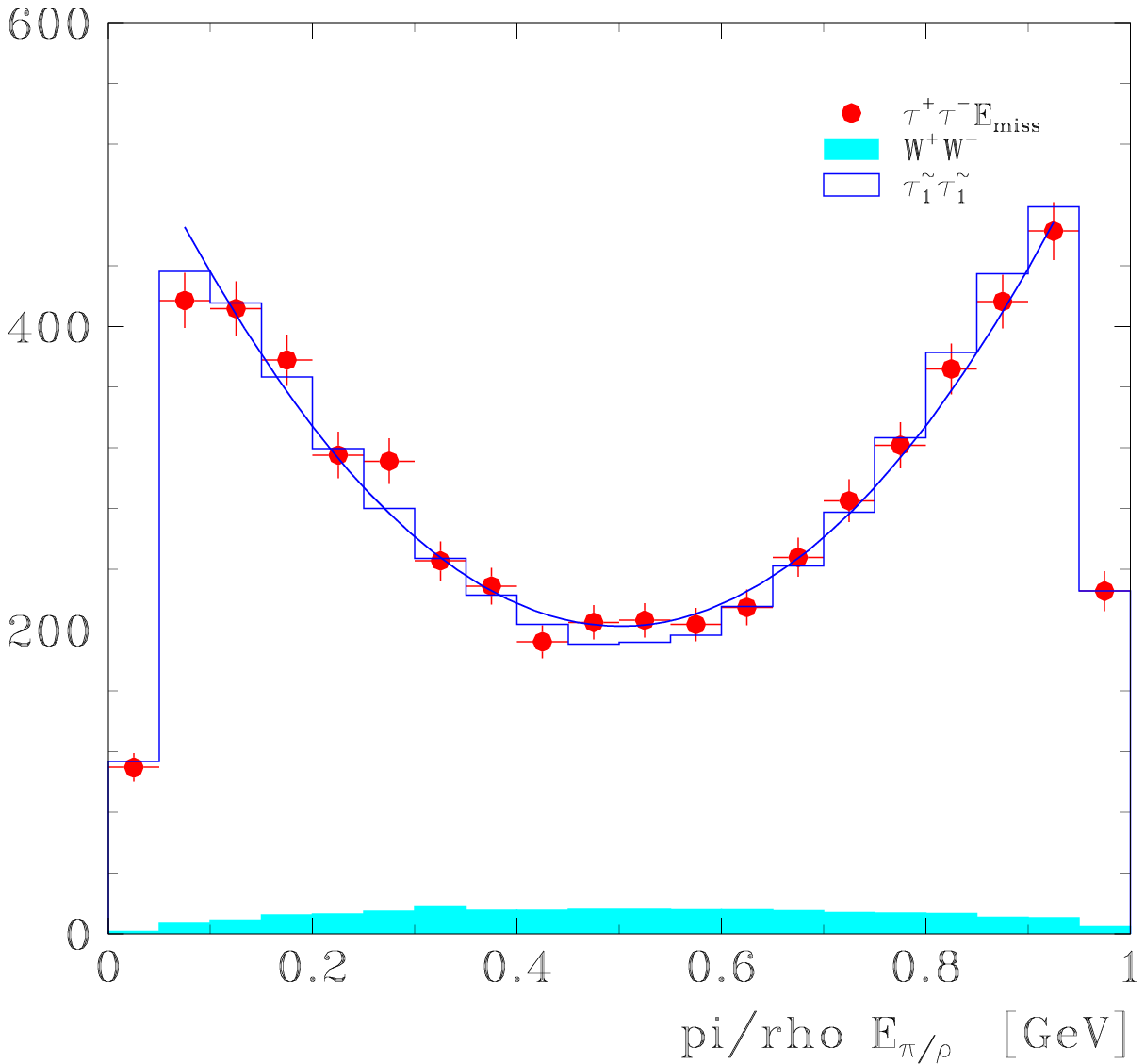,angle=90,width=.48\textwidth} 
  \caption{Energy spectrum
    $z_\pi = E_\pi/E_\rho$ from 
    $\tau \to\rho\; \nu_\tau\to \pi\pi^0\; \nu_\tau$ 
    of the reaction $e^+_L e^-_R\to \stau^+_1 \stau^-_1$ 
    including background from $W^+W^-\to\tau^+ \nu\; \tau^- \nu$. 
    The data points represent a simulation assuming $\cL = 200\,\fbi$ at 
    $\sqrt{s}=280\,\GeV$, 
    the blue histogram corresponds to a very high statistics sample, 
    the blue curve is a fit to the data with a $\stau_1$ polarisation 
    $\cP_\tau = 0.64 \pm 0.035$.} 
  \label{taupol} 
\end{figure} 
 
\quad\\

{\bf In summary}, simulations of stau production $\ee\to \stau_i\stau_j$ 
and chargino production $\ee\to \cx^+_1\cx^-_1$ under realistic 
experimental conditions assuming $(\cP_{e^-},\cP_{e^+}) = (0.8, 
0.6)$ show that the stau, neutralino and chargino masses as well 
as the polarisation $\cP_{\stau_1\to \tau\nt_1}$ and mixing 
parameter $\tstau$ can be accurately determined with moderate 
integrated luminosity. The possible ILC measurements are 
summarized in Table~\ref{tab:measurements}. 
 
\begin{table}[htb] 
\begin{center} 
\begin{tabular}{| l | l |} 
\hline\hline 
channel & observables \\ 
\hline 
      $\stau_1^+\stau_1^-$ & $m_{\stau_1} = 100.92 \pm 0.40~\GeV$ 
                                               \qquad $m_{\nt_1} = 80.67 \pm 0.35~\GeV$ \\ 
                                           & $\cos2\,\tstau = -0.065 \pm 0.028 $ 
                                               \qquad $\cP_\tau = 0.64 \pm 0.035 $  \\ 
      $\stau_2^+\stau_2^-$ & $m_{\stau_2} = 176.9 \pm 9.1~\GeV$  \\ 
      $\cx^+_1\cx^-_1$ & $m_{\cx^\pm_1} = 164.88 \pm 0.015~\GeV$  \\ 
\hline\hline 
\end{tabular} 
\end{center} 
\vspace*{-4mm}
\caption{Summary of achievable precisions of ILC measurements.} 
\label{tab:measurements} 
\end{table}

}
%---------------------------------------------------------------------------------------------------------
%---------------------------------------------------------------------------------------------------------

%---------------------------------------------------------------------------------------------------------
\section{DM properties: f\/it to ILC observables}
%---------------------------------------------------------------------------------------------------------

Having established how precisely one could measure various CP-even 
observables at ILC, we now estimate how this would constrain the
parameters of the underlying model, and the properties of the dark
matter candidate. To this aim we perform a fit to the six
observables listed in Table~\ref{tab:measurements} assuming
Gaussian errors. We also compute the total $\tau\tau$ SUSY cross
section at $\sqrt{s}=400$~GeV with polarized beams and add it to
the fit: $\sigma(\tau\tau)=3.220\pm 0.046$~pb (for 10~fb$^{-1}$).
Here we assume a systematic error as large as the statistical
error. As free parameters we take
\begin{equation}
  M_1,\quad \mu, \quad \tan\beta,  \quad
  M_{\tilde L_3}, \quad M_{\tilde R_3}, \quad A_\tau,
  \quad \phi_1, \quad \phi_\tau\,.
  \label{eq:freepar}
\end{equation}
Owing to the extremely small experimental error on $\mch{1}$, we
do not include $M_2$ as an independent variable in the fit but
compute it from $\mu$ and $\tan\beta$ such that it matches the
measured value of $\mch{1}$. The other SUSY parameters are fixed
to the values specified in \sect{benchmark}. We do not include the
light Higgs mass in the fit because of the large parametric
uncertainty from not knowing the parameters of the stop sector.
%The importance of the Higgs sector
This will be discussed in more detail after having summarized our main results.

To probe efficiently the multidimensional parameter space, we perform a Markov
Chain~\cite{mcmc} Monte Carlo (MCMC) analysis using a Metropolis-Hastings
algorithm~\cite{metro1,metro2,metro3}.
This algorithm generates a candidate state $x^c$ from the present state $x^t$ using a proposal
density $Q(x^t;x^c)$.  The candidate state is accepted to be the next state $x^{t+1}$ in the chain
if
\begin{equation}
  p = \frac{P(x^c)Q(x^t;x^c)}{P(x^t)Q(x^c;x^t)} \,,
\label{eq:pratio}
\end{equation}
where $P(x)$ is the probability calculated for the state $x$, is greater then a uniform random
number $a = U(0,1)$.  If the candidate is not accepted, the present state $x^t$ is retained and a new
candidate state is generated.  For the proposal density, we use a Gaussian distribution that is centered
at $x^t$ and has a width $\sigma$ for each parameter of  \eq{freepar}.
Moreover, we assume flat priors and take $P(x)=e^{-\chi^2(x)/2}$. A parameter point is hence accepted
in the Markov Chain, $x^{t+1}=x^c$, if
\begin{equation}
  p = e^{(\chi^2(x^{t}) - \chi^2(x^c))/2} > U(0,1).
\end{equation}
The $\chi^2$ function is computed as
\begin{equation}
  \chi^2(x) = \sum_i \frac{X_i(x)-\overline{X}_i}{\sigma(\overline{X}_i)}
\end{equation}
where $\overline{X}_i$ and $\sigma(\overline{X}_i)$ are the nominal values of the observables
and their 1-$\sigma$ errors, and $X_i(x)$ are the corresponding values obtained at the
parameter point $x$.
Note that, including the $\tau\tau$ cross section, we have seven measurements ($i=1,...,7$)
and eight free parameters, cf.\ \eq{freepar}.

%%%%%%%%%%%%%%%%%%%%%%%%%%%%%%%%%%%%%%%%%%%%%%%%%%%%
% chi2 vs mu and tan beta vs mu
\begin{figure}[t]
  \centerline{\epsfig{file=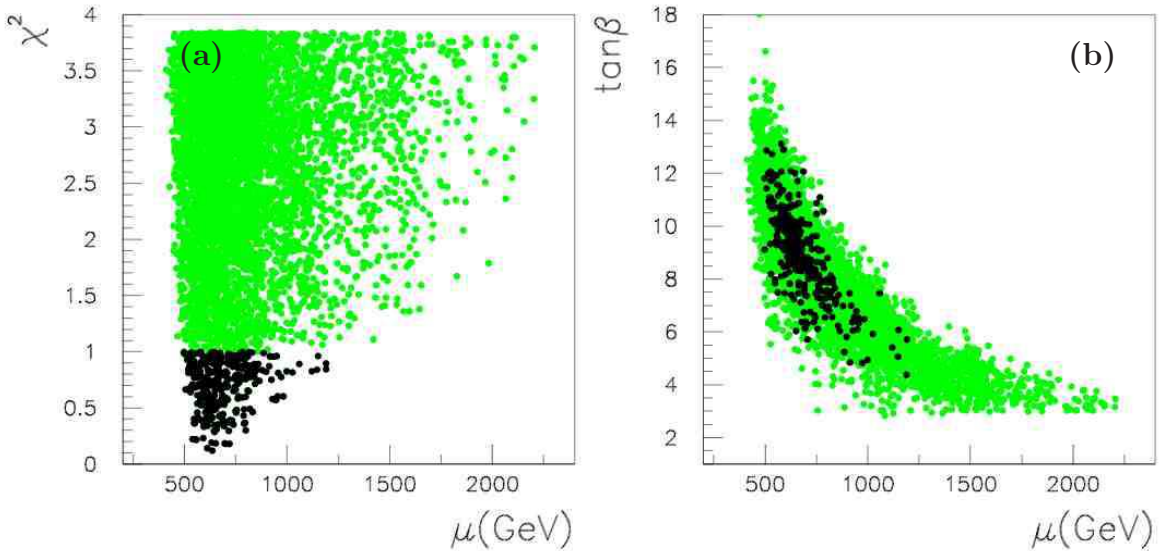, width=16cm}}
\vspace*{-6mm}
\ablabels{24}{141}{68}
  \caption{Results of the MCMC fit; in (a) $\chi^2$ as a function of $\mu$, in
  (b) the $1\sigma$ (black) and $2\sigma$ (green) allowed regions projected onto the
  $\mu$ versus $\tan\beta$ plane.}
\label{fig:fitres}
\end{figure}
%%%%%%%%%%%%%%%%%%%%%%%%%%%%%%%%%%%%%%%%%%%%%%%%%%%%

\begin{figure}[t]
  \centerline{\epsfig{file=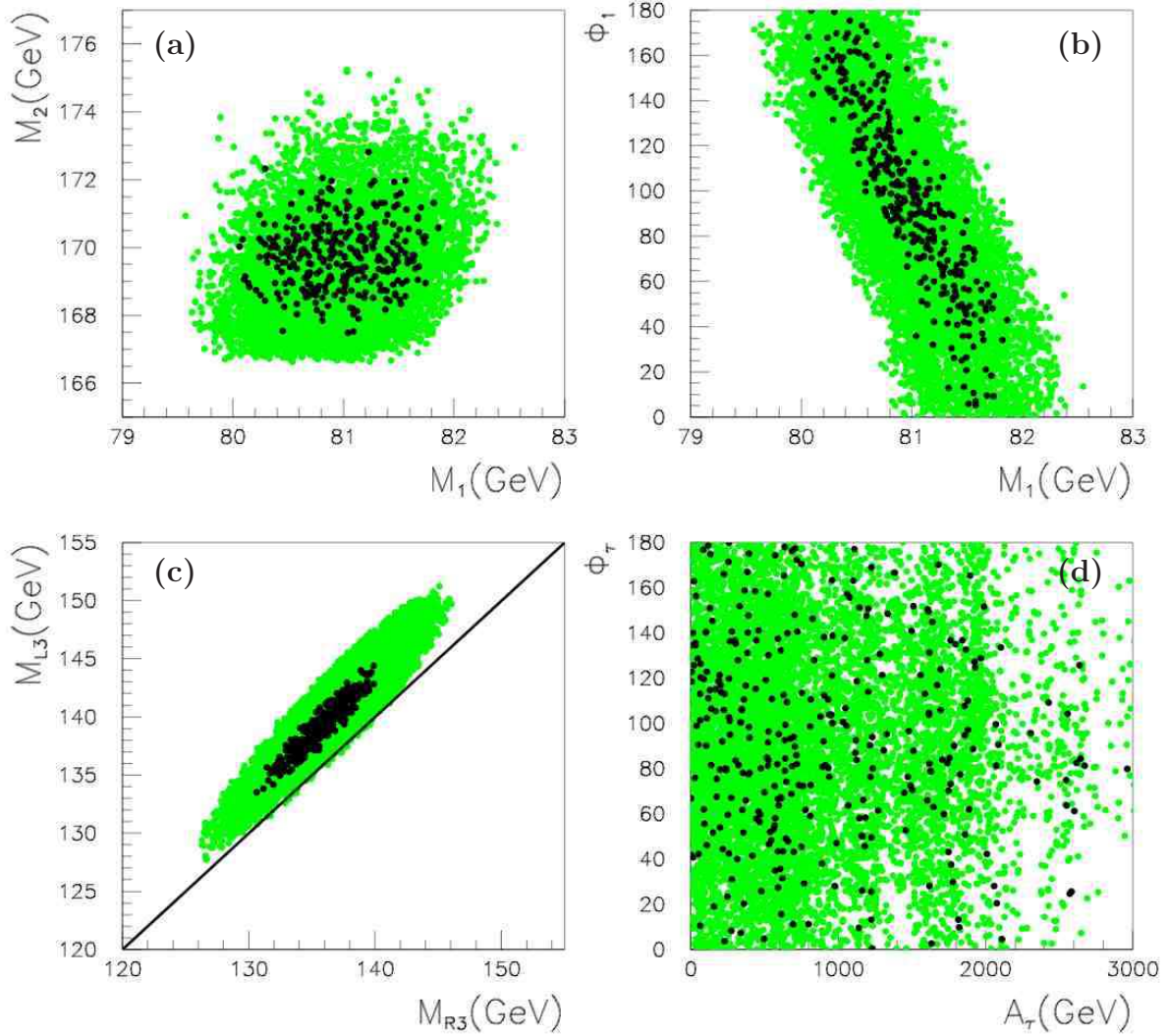, width=16cm}}
  \ablabels{21}{142}{142}
  \cdlabels{21}{142}{68}
  \vspace*{-5mm}
  \caption{The $1\sigma$ (black) and $2\sigma$ (green) allowed regions projected in different planes:
  (a)~$M_1$ vs.\ $M_2$, (b)~$M_1$ vs.\ $\phi_1$, (c)~$M_{\tilde R_3}$ vs.\ $M_{\tilde L_3}$
  and (d)~$A_\tau$ vs.\ $\phi_\tau$.}
\label{fig:param}
\end{figure}

We run $5\times 10^5$ points on 10 independent chains, with the starting points
having very different characteristics, such as large or small $\mu$; or
$M_{\tilde L_3}$ smaller, larger, or equal $M_{\tilde R_3}$.
We also include two randomly chosen starting points, for which we run $10^6$ points.
While some of the chains start off with huge $\chi^2$'s, they all converge
fast to $\chi^2\sim {\cal O}(1)$.
We do not only keep the points accepted in the chains but also write
out to a separate file all points tried which have $\chi^2< 3.84$,
corresponding to $95\%$\,CL.
%For these points we compute $\Omega h^2$, the Thallium EDM $d_{\rm Tl}$,
%as well as the spin-independent and spin-dependent direct detection cross sections
%$\sigma^{SI}_p$ and $\sigma^{SD}_p $.
\Fig{fitres}a shows the $\chi^2$ distribution of these points as a
function of $\mu$. It has a minimum around $\mu\sim 600$~GeV,
rises rather steeply for lower values and only slowly for higher
values  of $\mu$. We can gather from this plot that $\mu$ is
determined only to $500~{\rm GeV}\lsim\mu\lsim1200~{\rm GeV}$ at $1\sigma$
from the ILC measurements. The reason is a kind of `valley' of
$\chi^2$  minima in the $\tan\beta$--$\mu$ plane.  This is
illustrated in \fig{fitres}b which shows in green (black) points
which fall within the $2\sigma$ ($1\sigma$) region in the plane of
$\tan\beta$ versus $\mu$. As can be seen, while $\tan\beta$ and
$\mu$ are not very much constrained by the ILC measurements, there
is a strong correlation between the two parameters. Projections of
the $1\sigma$ and $2\sigma$ fitted regions onto the other
parameters are shown in \fig{param}. The fit results at 95\% CL
can be summarized as follows.

%%%%%%%%%%%%%%%%%%%%%%%%%%%%%%%%%%%%%%%%%%%%%%%%%%%%
% various parameter planes
%%%%%%%%%%%%%%%%%%%%%%%%%%%%%%%%%%%%%%%%%%%%%%%%%%%%

\begin{itemize}
\item{} The lack of observables individually sensitive to $\mu$
and $\tan\beta$ leaves a large allowed parameter space with
$400<\mu<2200$ and $2.8< \tan\beta <18$. Note, however, that because
the stau mixing is proportional to $(A_\tau-\mu\tan\beta)$ and because in
general $\mu\tan\beta\gg A_\tau$  these two parameters are strongly
correlated.

\item{} The very precise determination of $m_{\nt_1}$ and
$m_{\ch_1}$ constrains $M_1$ and $M_2$ to $\delta
M_1=-0.9,\,+2.1$~GeV and $\delta M_2=-3.7,\,+4.9$~GeV.

\item{} Although the phase $\phi_1$ can take any value, there is a
correlation with the value of $M_1$: a large phase is associated
with  lower values of $M_1$. This is a direct consequence of the
precise measurement of the neutralino mass.

\item{} The measurement of the masses of both staus and their mixing angle
 well constrain $M_{\tilde L_3}$ and $M_{\tilde R_3}$ to about 10~GeV.
 Moreover, solutions with $M_{\tilde L_3}<M_{\tilde R_3}$ are strongly disfavoured.

\item{} The trilinear coupling $A_\tau$ is basically undetermined, 
and the phase is $\phi_\tau$ completely unconstrained. 
This is because  the stau mixing is dominated by the term $\mu\tan\beta$.
The few points at low $\tan\beta\sim 4$ and $\mu\lsim 1$~TeV in \fig{fitres}{b}
have very large $|A_\tau|\lsim 3$~TeV. 
Setting an upper limit on $|A_\tau|$ somewhat tightens the lower border 
of the $2\sigma$ $\tan\beta$--$\mu$ range in \fig{fitres}b.

\end{itemize}

\begin{figure}
  \centerline{\epsfig{file=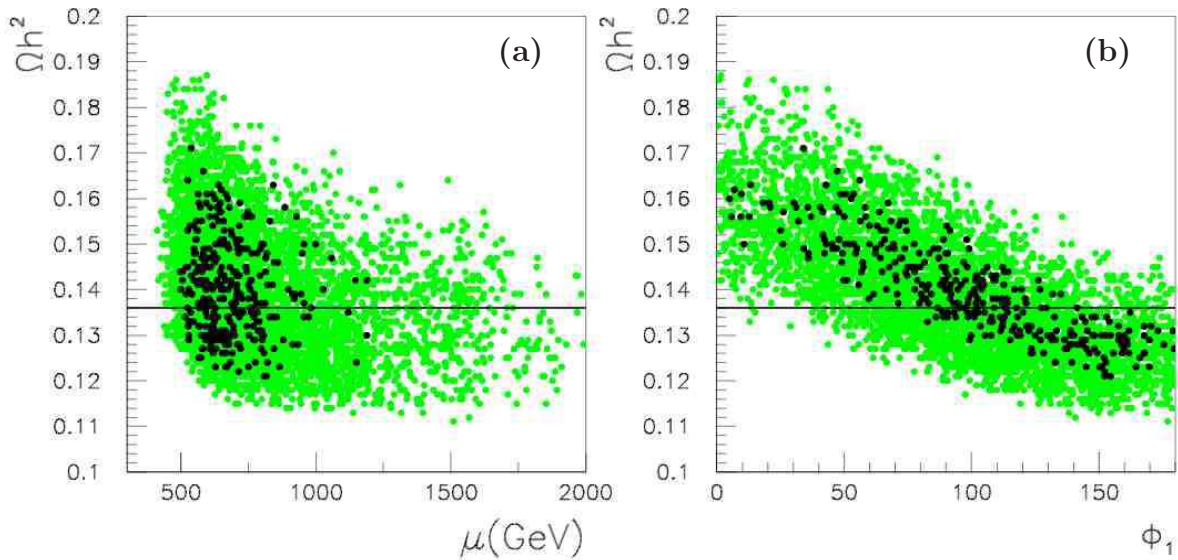, width=16cm}}
  \ablabels{66}{143}{75} \vspace*{-5mm}
  \caption{Predictions for (a) $\Omega h^2$ vs.\ $\mu$ and (b) $\Omega h^2$ vs.\ $\phi_1$
  in the $1\sigma$ (black) and $2\sigma$ (green) fitted regions. Straight lines show
  $2\sigma$ upper limits on $\Omega h^2=0.136$} \label{fig:omega}
\end{figure}

Having determined the allowed parameter space, we next show in
\fig{omega} the predictions for  the relic density of dark matter.
The allowed range at $2\sigma$ is found to be $0.116<\Omega
h^2<0.19$. As expected, $\Omega h^2$ tends to be larger than the
range that best fits present cosmological data. The lowest values
of $\Omega h^2$ are obtained for a large phase $\phi_1$ and both
$m_{\staul}$ and $m_{\staur}$ near their maximal allowed value.
While there is no strong correlation between $\mu$ and $\Omega h^2$, 
see Fig.~\ref{fig:omega}(a), it is only for $\mu\approx
500$--$700$~GeV that $\Omega h^2\gsim 0.18$. This is because for
small $\mu$ an increased higgsino component (although the LSP is
always dominantly bino) leads to a smaller annihilation
cross section into tau pairs. Furthermore, the mass of the LSP is
too small for this to be compensated by the very efficient
annihilation of the higgsino component into W pairs.

%%%%%%%%%%%%%%%%%%%%%%%%%%%%%%%%%%%%%%%%%%%%%%%%%%%%
% Omega vs mu and Omega vs phi_1
%%%%%%%%%%%%%%%%%%%%%%%%%%%%%%%%%%%%%%%%%%%%%%%%%%%%

\begin{figure}
  \centerline{\epsfig{file=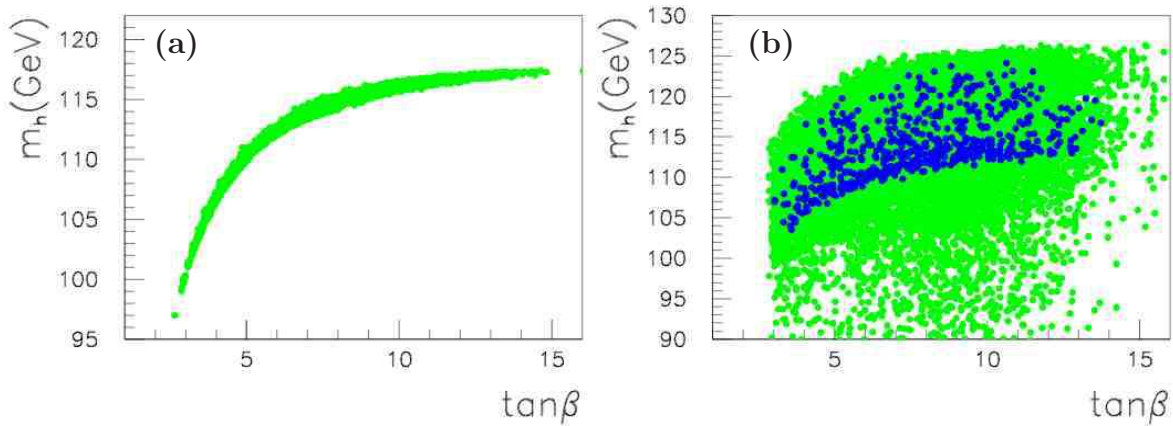, width=16cm}}
  \ablabels{21}{99}{59}\vspace*{-5mm}
  \caption{The $2\sigma$ region for $m_h$ as a function of $\tan\beta$: in (a) for parameters
  of the stop fixed to their nominal values and in (b) for varying stop parameters. In (b), the green
  points show the situation for a completely undetermined stop sector while the blue points are for
  stop masses known to $\pm 10\%$. }
\label{fig:mhstop}
\end{figure}

As mentioned above we do not include the light Higgs mass in our
fit. This may seem surprising since $m_h$ can be measured very
precisely at the ILC. Moreover, it strongly depends on $\tan\beta$
and might therefore be used to determine $\tan\beta$ and, in turn,
constrain $\mu$. However, one must not forget that $m_h$ also
strongly depends on the parameters of the stop sector. Indeed, if
the stop parameters are not known, this induces a parametric
uncertainty of about 15 GeV in $m_h$. As a consequence, the very
precise measurement of $m_h$ expected at the ILC only poorly
constrains $\tan\beta$ if the stop sector is not measured well.
For illustration, \fig{mhstop}(a) shows the correlation between
$m_h$ and $\tan\beta$ in the $2\sigma$ fitted region for stop
parameters fixed to their nominal values. \Fig{mhstop}(b) shows
the same but assuming that the stop masses are known to 10\%
(blue points) or that the stop sector is basically unknown (green
points). Here note that in our scenario even the discovery of the
stops is supposed to be very difficult at the LHC because of the
overwhelming $t\bar t$ background. A measurement of
$m_h=116.1$~GeV and stop masses known to 10\% accuracy would
constrain $\tan\beta$ only to about 3--14, theoretical
uncertainties not included. In turn, since in our fit the stop parameters
are fixed to 1 TeV, we allow Higgs masses as low as 100 GeV,
because one can always find a combination of $m_{\tilde t_1}$,
$m_{\tilde t_2}$ and $A_t$ which lifts $m_h$ above the LEP limit.

%%%%%%%%%%%%%%%%%%%%%%%%%%%%%%%%%%%%%%%%%%%%%%%%%%%%
% higgs mass versus tan beta
%%%%%%%%%%%%%%%%%%%%%%%%%%%%%%%%%%%%%%%%%%%%%%%%%%%%

%---------------------------------------------------------------------------------------------------------
\section{CP-odd observables}
%---------------------------------------------------------------------------------------------------------

In the previous sections, we have demonstrated how the parameters
$M_1$,  $M_2$, $\mu$, etc, of our stau-bulk scenario can be determined or at least 
constrained from an analysis of  particle masses, cross sections, distributions
at the ILC. Being CP even, such quantities principally allow to determine
CPV phases only up to a twofold ambiguity, $\phi \leftrightarrow 2\pi-\phi$.
In our scenario, only a correlation between $M_1$ and $\phi_1$ can be obtained, 
c.f. \fig{param}(b). In order to test for non-trivial phases, one needs 
CP-odd observables. These could be T-odd observables based on triple-products
in the production and decay of 
neutralinos or charginos, which are kinematically accessible in our scenario,
$e^+e^-\to\tilde\chi_1^0\tilde\chi_2^0$ and
$e^+e^-\to\tilde\chi_1^+\tilde\chi_1^-$, 
respectively.\footnote{
Note that there cannot be defined any appropriate T-odd asymmetries based
on triple products in
the pair production of staus, since they are scalar particles.}

%---------------------------------------------------------------------------------------------------------
\subsection{Neutralino production and decay}
%---------------------------------------------------------------------------------------------------------lala

\begin{figure}
  \centerline{\epsfig{file=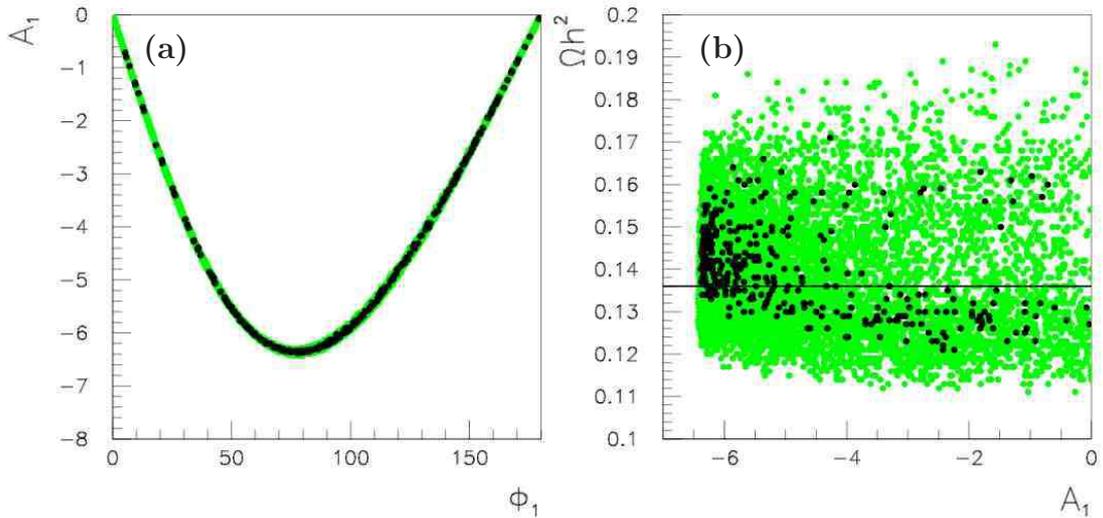, width=15cm}}
  \ablabels{25}{98}{70}\vspace*{-6mm}
  \caption{Triple-product asymmetry $A_1$ in the $1\sigma$ (black) 
  and $2\sigma$ (green) fitted regions,
  in (a) as a function of $\phi_1$ and in (b) correlation with $\Omega h^2$. 
%  {\bf OK: $A_1$ can only resolve ambiguities between the intervals
%    $0<\phi_1<\pi$ or $\pi<\phi_1<2\pi$, not in $0<\phi_1<\pi$ itself.}
}
\label{fig:asym}
\end{figure}

For the cross section $\sigma$ of neutralino production, 
$e^+e^-\to\tilde\chi_1^0\tilde\chi_2^0$,
followed by the subsequent two-body decay chain
$\tilde\chi_2^0\to\tilde\tau_1^\pm\tau^\mp$,
$\tilde\tau_1^\pm\to\tilde\chi_1^0\tau^\pm$,
we define the T-odd asymmetry~\cite{Kizukuri:1990iy,Bartl:2003tr,AguilarSaavedra:2004dz}
\begin{eqnarray}\label{A1}
  A_1 = 
      \frac{\sigma({\mathcal T}>0)
           -\sigma({\mathcal T}<0)}
           {\sigma({\mathcal T}>0)+
            \sigma({\mathcal T}<0)},
\end{eqnarray}
of the triple product 
\begin{eqnarray}\label{T1}
{\mathcal T} &=& ({\bf p}_{e^-} \times {\bf p}_{\tau^-}) 
      \cdot {\bf p}_{\tau^+},
\end{eqnarray}
of the three-momenta ${\bf p}$.
In \Fig{asym}(a), we show the triple-product asymmetry $A_1$ 
as a function of $\phi_1$ in the $1\sigma$ and $2\sigma$ fitted regions. 
As can be seen, a measurement of 
$A_1\not=0$ would be a very clear signal 
of CP violation.\footnote{The asymmetry $A_1$ flips sign for $\phi_1\in (\pi,2\pi)$, 
while the CP-even observables, including $\Omega h^2$, are symmetric around $\pi$.} 
Unfortunately, this does not help constrain 
$\Omega h^2$ due to the two-fold 
ambiguity in $\phi_1$, see \fig{asym}(b).
Note also that in our scenario, the neutralino production cross sections
are suppressed due to the heavy selectron masses $m_{\tilde e_{L,R}}=1$~TeV,
however can be as large as $10$~fb for $(P_{e^-},P_{e^+})=(0.8,-0.6)$
at $\sqrt s = 500$~GeV. 
The branching ratios are of the order of
${\rm BR}(\tilde\chi_2^0\to\tilde\tau_1^\pm\tau^\mp)=50\%$.

%%%%%%%%%%%%%%%%%%%%%%%%%%%%%%%%%%%%%%%%%%%%%%%%%%%%
% triple-prod asym
%%%%%%%%%%%%%%%%%%%%%%%%%%%%%%%%%%%%%%%%%%%%%%%%%%%%

In principle, the \emph{transverse} polarisation of the $\tau$
from the decay $\tilde\chi_2^0\to\tilde\tau_1\tau$ is sensitive to
the phases $\phi_1,\phi_\mu,\phi_\tau$, and can be large in 
general~\cite{Bartl:2003gr, Choi:2003pq}.
However, since the  dependence of the transverse $\tau$ polarization
on $\phi_\tau$ is weak for $|A_\tau| \ll |\mu|\tan\beta$,
as in our scenario, we only obtain polarisations not larger than 
$0.7\%$ for non-vanishing phases, with
${\rm BR}(\tilde\chi_1^+\to\tilde\nu_\tau\tau^+)\approx50\%$.

%---------------------------------------------------------------------------------------------------------
\subsection{Note on chargino production}
%---------------------------------------------------------------------------------------------------------

CP-odd observables in chargino production and
decay are in general only sensitive to the phase $\phi_\mu$
of the chargino sector. Our scenario actually assumes
$\phi_\mu=0$. For completeness, we briefly mention 
how to constrain the phase $\phi_\mu$ in principle.
Due to the heavy sneutrino mass $m_{\tilde\nu_e}=1$~TeV, the 
destructive sneutrino interference term in chargino production
$e^+e^-\to\tilde\chi_1^+\tilde\chi_1^-$ is suppressed,
such that the cross section reaches more than
$2000$~fb in our scenario for $(P_{e^-},P_{e^+})=(-0.8,0.6)$
at $\sqrt s = 500$~GeV.
Since all couplings in  
diagonal chargino pair production are real~\cite{Choi:1998ei}, 
there is no CP-sensitive contribution
from $\phi_\mu$ in the production process alone.
However, if the chargino decays into polarised particles,
viz. the $\tau$, 
CP-sensitve observables can be defined~\cite{marold}.
Similar to the neutralino decay, one such quantity is the 
the transverse polarisation of the $\tau$ in the decay
$\tilde\chi_1^\pm\to\tau\tilde\nu_\tau$.
In our scenario, the transverse $\tau$ polarisation
could be as large as $4\%$ for $\phi_\mu=90^\circ$.

%--------------------------------------------------------------------------------------------------------------------------
\section{Constraining further the model}
%--------------------------------------------------------------------------------------------------------------------------

In the previous section we have found that the precise
determination of a few observables at the ILC still leaves a large
uncertainty in the prediction of the relic density of the dark
matter candidate. This holds even if CP-odd observables are included.
However, at the timescale where the ILC may start
its operation, other measurements will be available that could
constrain further the model. 
Below we discuss direct and indirect detection of DM, as well as 
EDM measurements. 
Finally, we comment on further measurements of heavy sparticles at 
the LHC and/or a linear collider with higher energies.

\subsection{Direct detection}

Upper limits  on the cross section for elastic scattering of DM particles on nuclei are 
improving every year, with the strongest constraints currently coming from 
Xenon10~\cite{Angle:2007uj} and CDMS~\cite{Ahmed:2008eu}, see also \cite{dmtools}.  
For a $80$~GeV WIMP, the limit on the spin-independent interaction with protons is 
$\sigma^{SI}_p\lsim 5\times 10^{-8}$~pb~\cite{Ahmed:2008eu}.
The next generation of detectors should probe $\sigma^{SI}_p\approx 10^{-10}$~pb 
within less than 10 years~\cite{Kraus:2007zz}. 
Furthermore, the near maximal sensitivities of the
detectors are in the mass range of interest here, i.e.\ around
100~GeV. For spin-dependent interactions, limits are much weaker,
about $10^{-2}$~pb~\cite{Alner:2007xs,Akerib:2005za} with prospects 
of reaching ultimately $\sigma^{SD}_p\approx 10^{-6}$--$10^{-7}$~pb  
with ton-size detectors, see e.g.~\cite{Collar:2007xn}.  

%%%%%%%%%%%%%%%%%%%%%%%%%%%%%%%%%%%%%%%%%%%%%%%%%%%%
% direct detection cross sections
\begin{figure}
  \centerline{\epsfig{file=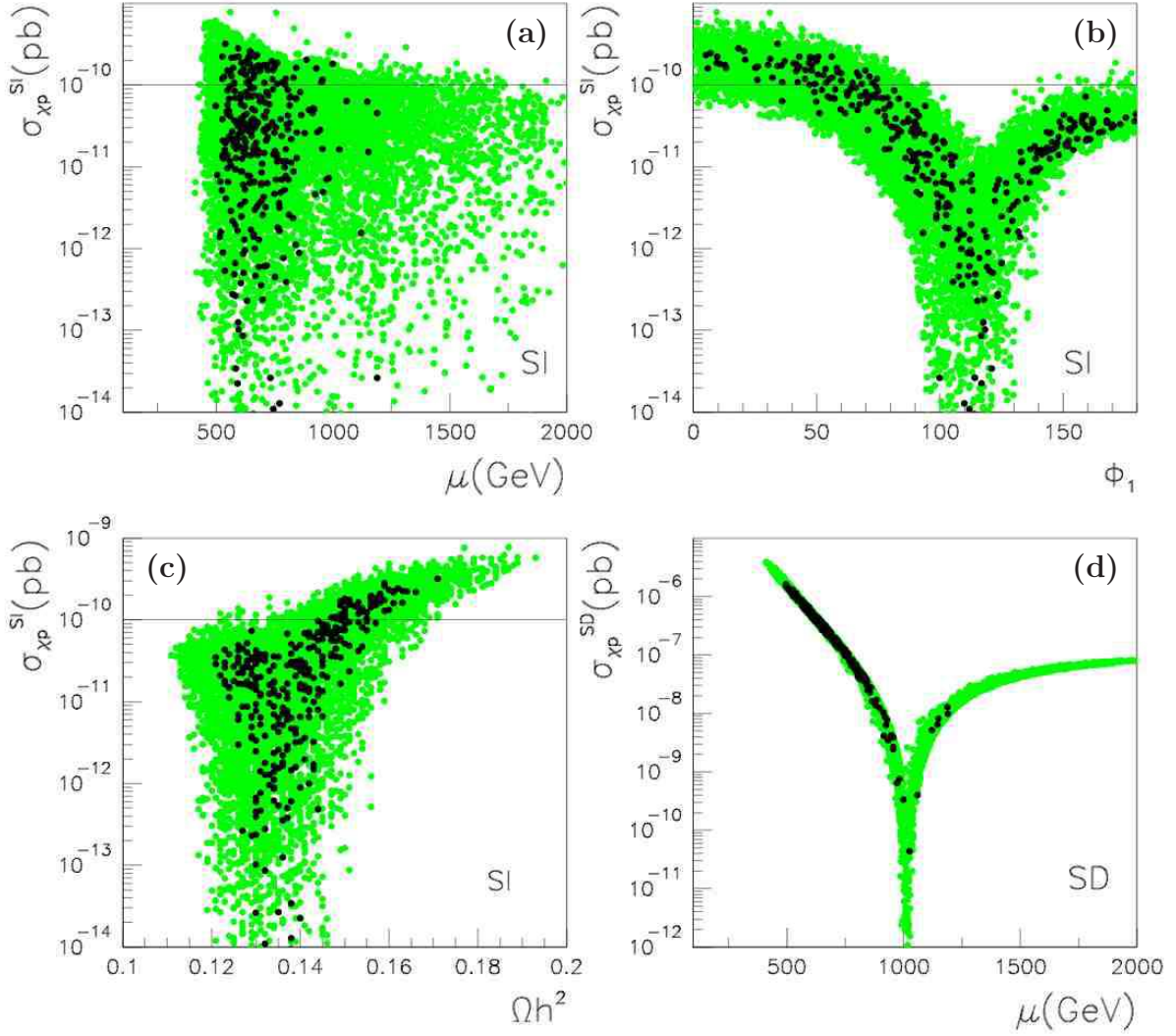, width=16cm}}
  \ablabels{68}{144}{143}
  \cdlabels{20}{144}{68}
  \vspace*{-4mm}
  \caption{Predictions for $\sigma^{SI}_p$, in (a) as a function of $\mu$,
  in (b) as a function of $\phi_1$. In (c) correlation between $\sigma^{SI}_p$
  and $\Omega h^2$.  The expected reach of ton-size detectors is indicated.
  Frame (d) shown predictions for $\sigma^{SD}_p$  as a function of $\mu$.
  As before, the $2\sigma$ ($1\sigma$) fitted region is shown in green (black).}
\label{fig:dirxsect}
\end{figure}
%%%%%%%%%%%%%%%%%%%%%%%%%%%%%%%%%%%%%%%%%%%%%%%%%%%%

In our scenario, where the squarks of the first two generations
are heavy, the spin-independent cross section is completely
dominated by the Higgs exchange diagram. We therefore expect
$\sigma_p^{SI}$ to be sensitive to $\mu$ and $\phi_{1}$ through
the $\nt_1\nt_1 h$ coupling. Since the LSP couples to the Higgs
through its higgsino component, we expect a small cross section in
our scenario, where the LSP is dominantly bino.  This is indeed the case,
as can be seen in \fig{dirxsect}. Nevertheless for some of the
parameter space the cross section is large enough  to be
detectable at future ton-size detectors. This occurs mostly for near
minimal $\mu$ values, \fig{dirxsect}(a). The dependence on the phase
$\phi_1$, \fig{dirxsect}(b), can be related to the suppression of
the $\nt_1\nt_1 h$ coupling. Furthermore, for near maximal phase
there is an interference between the light and heavy Higgs
contributions which leads to a strong suppression of the $\nt_1
p\rightarrow \nt_1 p$ cross section. The models which predict a
cross section that could be detected in the near future are
precisely those where $\Omega h^2$ is the largest, see 
\fig{dirxsect}(c). Since a positive signal is not expected in this
particular scenario, we conclude that an improvement on the limit
of the neutralino proton elastic scattering would reduce the
uncertainty in the prediction of the relic density to 
$0.116\le\Omega h^2\le 0.17$, provided the major hadronic uncertainties
\cite{Bottino:2001dj,Ellis:2008hf} stemming from the strange content in the nucleon can be 
brought under control.

The spin-dependent cross section is dominated by  $Z$ exchange and
hence is also expected to be largest when the higgsino component is
largest, that is for small $\mu$. No strong dependence on
the phase $\phi_1$ appears. 
Moreover, as shown in \fig{dirxsect}(d), the prediction for the spin-dependent 
cross section is far below the present limit. The $\mu$-dependence is, however, 
more pronounced than in the spin-independent case. 
If ambitious projects like \cite{Collar:2007xn} can reach $\sim10^{-7}$~pb, they 
will indeed probe the small $\mu$ region, where our nominal point lies.

Last but not least note that the cross sections for direct DM detection also depend on 
the local DM density and velocity distribution. Therefore even a positive DM signal 
will not directly limit the SUSY parameter space, but rather provide a consistency check.

\subsection{Cross sections for indirect detection}

\begin{figure}
  \centerline{\epsfig{file=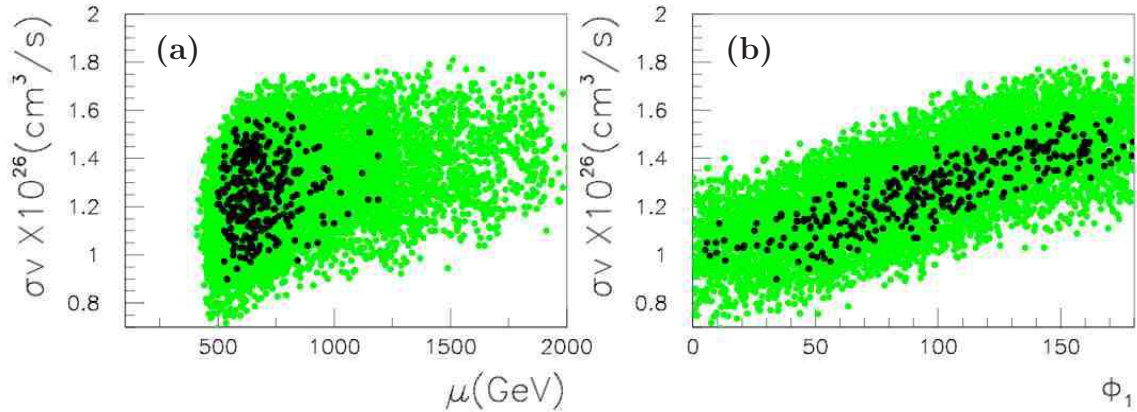, width=15.5cm}}
  \ablabels{23}{98}{56}
  \vspace*{-6mm}
  \caption{Indirect detection cross sections $\sigma v$ [$cm^3/s$]
  in the $1\sigma$ (black) and $2\sigma$ (green) regions,
  in (a) as a function of $\mu$, and in (b) as a function of $\phi_1$.}
\label{fig:indirect}
\end{figure}

For completeness we show in \fig{indirect} also the cross section for neutralino annihilation into
photons. Here we consider only photons that come from the decays of the neutralino
annihilation products.
The cross-section is of the order of few times $10^{-26}~{\rm cm^3/s}$ and vary only by about a factor of 
two in the $2\sigma$ parameter range. 
These cross sections are relevant for indirect DM detection experiments such as 
EGRET~\cite{Sreekumar:1997yg} or GLAST
~\cite{Morselli:2007zza}. 
While of course being 
interesting and useful by itself, indirect detection will not help in pinning down 
the model parameters. Indeed the rate for $\gamma$-rays detection strongly depends  on the
model for the neutralino density near the center of the galaxy, predictions can vary by more  
than an order of magnitude~\cite{Bertone:2004pz} washing out the  variations induced by the
phase dependence.

%%%%%%%%%%%%%%%%%%%%%%%%%%%%%%%%%%%%%%%%%%%%%%%%%%%%
% indirect detection 
%%%%%%%%%%%%%%%%%%%%%%%%%%%%%%%%%%%%%%%%%%%%%%%%%%%%

\subsection{EDMs}

\begin{figure}
  \centerline{\epsfig{file=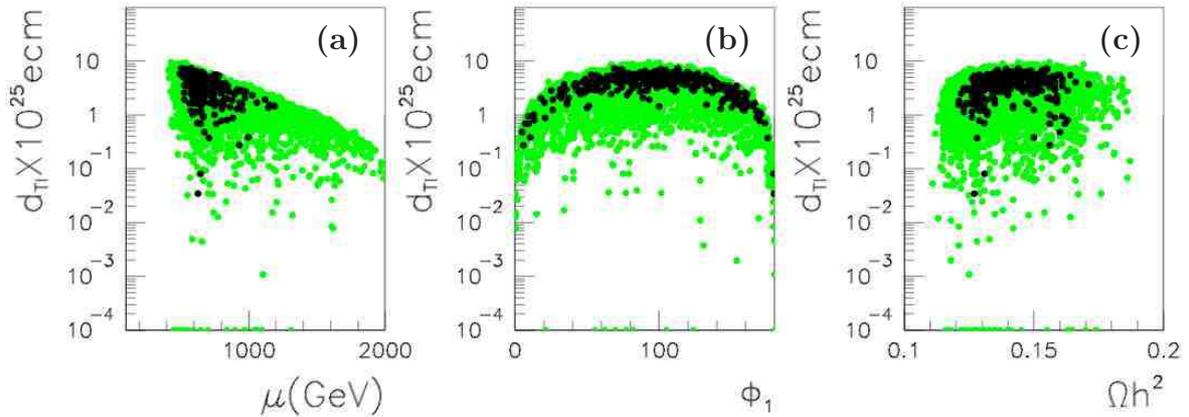, width=16cm}}
  \begin{picture}(100,0)\setlength{\unitlength}{1mm}
    \put(42,59){\bf (a)}
    \put(93,59){\bf (b)}
    \put(145,59){\bf (c)}
  \end{picture}
  \vspace*{-9mm}
  \caption{Predictions for $d_{\rm Tl}$ in the $1\sigma$ (black) and $2\sigma$ (green) regions, 
  in (a) as a function of $\mu$ and (b) as a function of $\phi_1$.
  Frame (c) shows the correlation between $d_{\rm Tl}$ and $\Omega h^2$.}
\label{fig:edm}
\end{figure}

We argued in \sect{benchmark} that EDM constraints are satisfied
in our scenario by choosing heavy sfermions of the first and second
generations. However, scanning over the
parameters of the model we find that for not too large values of
$\mu$ and a large phase $\phi_1$, the predictions for the Thallium
EDM are above the present bound
$d_{\rm Tl}<9 \times 10^{-25}~{\rm e\,cm}$~\cite{Regan:2002ta}, see Fig.~\ref{fig:edm}(a,b).
Taking this bound into account would, however, not much impact the
allowed range for $\Omega h^2$ as can be seen in Fig.~\ref{fig:edm}(c). 
Note that here $d_{\rm Tl}$  is computed using $M_{\tilde e,\tilde\mu}\simeq1$~TeV 
and that a lack of discovery of sleptons at LHC/ILC will only put a lower bound on the 
slepton mass. Increasing the slepton masses beyond 1~TeV further weakens the EDM
constraint. For this reason we do not include this contraint
in our fit but use it only {\it a posteriori}. The electron EDM is currently
directly extracted from the Thallium measurement, prospects for improving the
accuracy on the electron EDM by two orders of magnitude are being 
explored~\cite{Kawall:2003ga,Lamoreaux:2007rt}.
Clearly such refined measurements should probe most of the parameter space of the
scenario considered.

%%%%%%%%%%%%%%%%%%%%%%%%%%%%%%%%%%%%%%%%%%%%%%%%%%%%%%
% edm plots
%%%%%%%%%%%%%%%%%%%%%%%%%%%%%%%%%%%%%%%%%%%%%%%%%%%%%%

\subsection{Heavy particles at colliders}\label{sec:lhc}

A large part of the uncertainty in the model reconstruction comes from 
the lack of knowledge of $\mu$ and $\tan\beta$. Here the obvious ways 
out would be to i)~determine $\tan\beta$ from the Higgs sector and/or 
ii)~determine $\mu$ from a measurement of the higgsino states. 

As explained above, to exploit the precision measurement of the light Higgs 
mass, one would need to know the parameters of the stop sector. 
Extracting a stop signal is, however, notoriously difficult at the LHC,
at least unless $\st_1$ is very light \cite{Kraml:2005kb,Allanach:2006fy}. 
This is due to the sheer overwhelming $t\bar t$ background. 
Moreover, even if the stop masses could be measured, this would not 
be sufficient; one would need in addition a measurement of the 
stop mixing angle to extract $A_t$. This only seems feasible at 
a multi-TeV $e^+e^-$ linear collider, such as CLIC~\cite{Accomando:2004sz}. 

The heavy higgsino-like chargino $\ti\chi^\pm_2$ and neutralinos 
$\nt_{3,4}$ could be detected at the LHC through Drell-Yan production 
or through squark decays into them. The former suffers from a very 
small cross section of only few fb, and the latter from tiny branching ratios.
It is clear that any such 
measurement will be very challenging and require very high luminosity.  

Alternatively, $\ti\chi^\pm_2$ or $\nt_{3,4}$ could be produced in 
$e^+e^-$ annihilation with higher energy. 
The first kinematically accessible process is $\nt_1\nt_{3,4}$ at about 700 GeV, 
followed by $\ti\chi^\pm_1\ti\chi^\mp_2$ at about 800 GeV. 
However, the production cross sections are small, below a femtobarn, since 
one state is almost purely bino or wino and the other higgsino. 
To be precise, at $\sqrt{s}=800$~GeV one would have 
$\sigma(e^+_Le^-_R\to\ti\x^\pm_1\ti\x^\mp_2) = 0.74$~fb and
$\sigma(e^+_Le^-_R\to\nt_2\nt_3) = 0.45$~fb, 
increasing to $1.48$~fb and $0.74$~fb respectively at $\sqrt{s}=1$~TeV.
At $\sqrt{s}\ge1.22$~TeV, one could finally produce the heavier states in pairs, 
i.e.\ $\ti\chi^\pm_2\ti\chi^\mp_2$, $\nt_3\nt_3$, etc..
The $\ti\x^\pm_2\ti\x^\mp_2$ cross sections rises fast, giving 
$\sigma(e^+_Re^-_L\to\ti\x^+_2\ti\x^-_2)\simeq 100$~fb at $\sqrt{s}\ge1.4$~TeV, while 
$\sigma(e^+_Re^-_L\to\nt_3\nt_4)\simeq 30$~fb at this energy and the other cross 
sections much smaller. 
With such measurements, the complete neutralino/chargino system can 
in principle be reconstructed~\cite{Kneur:1999nx,Choi:2000ta,Choi:2001ww,Choi:2002mc}.

A detailed analysis of LHC measurements or linear collider measurements at TeV energies 
is beyond the scope of this paper. Notice, however, that for a percent-level collider determination 
of $\Omega h^2$, one would need to know not only $\mu$ and $\tan\beta$ but also the CP phases 
(in our case $\phi_1$) to a comparable, i.e. percent-level, precision.

\section{Conclusions}

Considering the general MSSM with CP phases, 
we have investigated a ``stau-bulk'' scenario, in which only
gauginos and staus are light and accessible to ILC measurements, 
while all other sparticles are heavy. In this scenario, 
a precise determination of SUSY particle properties is quite
challenging at colliders because of the dominance of the channels
involving taus and missing energy. Combining threshold scans,
endpoint methods, measurements of polarized cross section and tau
polarization, the masses of the lightest neutralino, chargino and
of both staus can be determined precisely at the ILC. Information
on the stau mixing can also be obtained. 

From these measurements, some of the underlying Lagrangian parameters 
can be extracted with good precision: $M_1$, $M_2$, $M_{\ti L_3}$,  $M_{\ti R_3}$. 
Moreover, the product $\mu\tan\beta$ can be constrained. However, we are 
left with a large uncertainty in both $\mu$ and $\tan\beta$, as well as in the 
phase of $M_1$, $\phi_1$. This causes a rather large uncertainty in the collider 
prediction of the neutralino relic density of $0.116<\Omega_{\ti\chi} h^2<0.19$ at 95\% CL. 

Taking into account the possibility of non-zero phases is of particular importance:  
from pure sparticle spectroscopy, e.g.\ the measurement of the $\nt_1$--$\stau_1$ 
mass difference which is $\sim 10$ GeV in our scenario, one might conclude that 
the neutralino relic density is too large, $\Omega_{\ti\chi} h^2\gsim 0.14$.
Evidence of a CP-violating signal in EDMs and/or in collider measurements 
would clearly show that phases have to be taken into account. 
It would, however, not directly add information for infering the relic density 
of the neutralino dark matter candidate.

We have also discussed implications for (in)direct dark matter searches. 
While the cross section for indirect detection shows a too weak dependence, 
information from large scale dark matter detectors could indeed somewhat 
reduce the allowed parameter range. In any case, direct and indirect detection 
are important cross checks that the $\ti\chi^0_1$ is indeed the DM.

Finally, for a percent-level collider determination of $\Omega h^2$, which matches
the precision of cosmological observations, one would need know also $\mu$, $\tan\beta$,
and the CP phases (in our case $\phi_1$) to percent-level precision.
To this aim, the above-mentioned ILC measurements have to be complemented by 
precision measurements of the heavy higgsino-like neutralinos and charginos 
at TeV energies.

%---------------------------------------------------------------------
\section*{Acknowledgements}
%---------------------------------------------------------------------
\addcontentsline{toc}{section}{Acknowledgements}

We thank S.~Sekmen and R.\,K.~Singh for useful discussions on the MCMC. 
O.K.\ thanks T.~Kernreiter, J.A.~Aguilar-Saavedra, and K.~Hohenwarter-Sodek 
for clarifying discussions on triple products.

This work was supported in part by GDRI-ACPP of CNRS,  
the `SFB Transregio 33: The Dark Universe', and grant 
RFBR-08-02-00856-a of the Russian Foundation for Basic Research. 
This work is also part of the French ANR project ToolsDMColl. 
A.P. acknowledges the hospitality of CERN and LAPTH where some of the
work contained here was performed.

%---------------------------------------------------------------------
%\bibliography{cp-ref}
%---------------------------------------------------------------------

%---------------------------------------------------------------------
\end{document}